\documentclass[preprint2]{aastex}

\newcommand{\etal}{et~al.}
\newcommand{\eg}{e.g., }
\newcommand{\ie}{i.e., }
\newcommand{\Msun}{M_{\odot}}

\newcommand{\Nifs}{$^{56}$Ni}

\newcommand{\Mms}{M_{\rm MS}}

\def\gsim{\mathrel{\rlap{\lower 4pt \hbox{\hskip 1pt $\sim$}}\raise 1pt
\hbox {$>$}}}
\def\lsim{\mathrel{\rlap{\lower 4pt \hbox{\hskip 1pt $\sim$}}\raise 1pt
\hbox {$<$}}}

\slugcomment{Accepted for publication in the Astrophysical Journal
Supplement Series (2007 May).}

\begin{document}

\title{Supernova Nucleosynthesis in Population III 13 -- 50 $\Msun$ Stars
and Abundance Patterns of Extremely Metal-Poor Stars}

\author{
 Nozomu~Tominaga\altaffilmark{1},
 Hideyuki~Umeda\altaffilmark{1}, and
 Ken'ichi~Nomoto\altaffilmark{1,2}
 }

\altaffiltext{1}{Department of Astronomy, School of Science,
University of Tokyo, Bunkyo-ku, Tokyo 113-0033, Japan;
tominaga@astron.s.u-tokyo.ac.jp, umeda@astron.s.u-tokyo.ac.jp, nomoto@astron.s.u-tokyo.ac.jp}
\altaffiltext{2}{Research Center for the Early Universe, School of
Science, University of Tokyo, Bunkyo-ku, Tokyo 113-0033, Japan}

\begin{abstract}
We perform hydrodynamical and nucleosynthesis calculations of
core-collapse supernovae (SNe) and hypernovae (HNe) of Population
(Pop) III stars. We provide new yields for the main-sequence mass of
 $\Mms=13-50$ $\Msun$ and the explosion energy of $E=1-40\times10^{51}$
 ergs to apply for chemical evolution studies.  Our HN yields based on
 the mixing-fallback model of explosions reproduce the 
observed abundance patterns of extremely metal-poor (EMP) stars
($-4< {\rm [Fe/H]}< -3$), while those of very metal-poor (VMP)
stars ($-3<{\rm [Fe/H]} < -2$) are reproduced by the
normal SN yields integrated over the Salpeter initial mass function.
Moreover, the observed trends of abundance ratios [X/Fe] against
[Fe/H] with small dispersions for the EMP stars can be reproduced as a
sequence resulting from the various combination of $\Mms$ and $E$.  This
 is because we adopt the
empirical relation that a larger amount of Fe is ejected by more
massive HNe.  Our results imply that the observed trends with small
dispersions do not necessarily mean the rapid homogeneous mixing in
the early galactic halo at [Fe/H] $< -3$, but can be reproduced by
the ``inhomogeneous'' chemical evolution model.
 In addition, we examine how the modifications of the distributions
 of the electron mole fraction $Y_{\rm e}$ and the density in the
 presupernova models improve the agreement with observations. In this
 connection, we discuss possible contributions of nucleosynthesis in the
 neutrino-driven wind and the accretion disk.
\end{abstract}

\keywords{Galaxy: halo 
--- nuclear reactions, nucleosynthesis, abundances 
--- stars: abundances --- stars: Population III 
--- supernovae: general
\begin{center}
{\Large {\bf Accepted for publication in \\
the Astrophysical Journal (10 May 2007, v660n2 issue).}}
\end{center}
}

\section{INTRODUCTION}
\label{sec:intro}

The first metal enrichment in the universe was made by the supernova
(SN) explosions of population (Pop) III stars. Despite the importance of
Pop III stars in the evolution of the early universe, their
properties are still uncovered. 
The main issues is the typical masses of Pop III stars. Some studies have
suggested that the initial mass function (IMF) differs from the present
day IMF (\eg top-heavy IMF; \citealt{nakf99,bro04}) and that a large
number of stars might be so massive as to explode as pair-instability SNe
(\eg \citealt{was00}). On the other hand, \cite{tum05} suggested an
IMF that is peaked in the range of massive stars that exploded as
core-collapse SNe. 

In the early universe, the enrichment by a single SN can dominate the
preexisting metal contents (\eg \citealt{aud95}).
The Pop III SN shock compresses the SN ejecta consisting of heavy
elements, \eg O, Mg, Si, and Fe, and the circumstellar materials
consisting of H and He, and thus the abundance pattern of the enriched gas may
reflect nucleosynthesis in the SN. The SN compression will initiate a
SN-induced star formation (\eg \citealt{cio88,rya96,shi98}) and the second-generation
stars will be formed from the enriched gas. Among the second generation
stars, low mass ($\sim 1\Msun$) stars have long life-times and might be
observed as extremely metal-poor (EMP) stars with [Fe/H] $< -3$
\citep{bee05}. (Here [A/B] 
$= \log_{10}(N_{\rm A}/N_{\rm B})-\log_{10} (N_{\rm A}/N_{\rm B})_\odot$, 
where the subscript $\odot$ refers to the solar value and $N_{\rm A}$
and $N_{\rm B}$ are the abundances of elements A and B, respectively.)
Therefore the EMP stars should conserve the nucleosynthetic
results of the Pop III SN and can constrain the yields of the SN. 

The elements ejected by various SNe are gradually mixed and the abundance
patterns of the Galaxy becomes homogeneous with time. The abundance
patterns of the newly formed stars reflect averaged nucleosynthesis
over various SNe. It is important to know when the transition
from inhomogeneous to homogeneous mixing occurs. The timing of this
transition can be informed from chemical evolution calculations with
hierarchical models; \cite{arg00} has suggested that a halo ISM is unmixed
and inhomogeneous at [Fe/H] $<-3.0$, intermediate between unmixed and
well mixed at $-3.0<$ [Fe/H] $<-2.0$, and well mixed at [Fe/H] $>-2.0$;
\cite{tum05} has suggested that the mean number of reflected SNe is 10
at ${\rm [Fe/H]} \sim -2.8$.

The previous observations (\eg McWilliam et al. 1995a,b; \citealt{rya96,mcw97})
provide the abundance patterns of the EMP stars that show interesting trends of
elemental abundance ratios [Cr/Fe], [Mn/Fe], [Co/Fe], [Zn/Fe] with
decreasing [Fe/H], although dispersions are rather large.
These trends, except for the absolute values of some elements, can be
explained by the differences of the progenitors' masses and the
explosion energies assuming the SN-induced star formation
(\citealt{nak99}; Umeda \& Nomoto 2002a, 2005, hereafter \citealt{ume02},
\citealt{ume05}).

Recent observations for $-4 \lsim {\rm [Fe/H]} \lsim -2$ by Cayrel \etal\
(2004, hereafter \citealt{cay04}) confirmed these trends shown by the
previous studies with much smaller dispersions (see, however Honda
\etal\ 2004, hereafter \citealt{hon04}, for the difference in [Cr/Fe] at 
$-3 \lsim {\rm [Fe/H]} \lsim -2$), except for much flatter trends of
[Mg/Fe] and [Mn/Fe] than the previous studies. \cite{cay04} and
\cite{fra04} suggested the following interpretation of the observed
small dispersions: the elements have been already mixed homogeneously in
the halo even below [Fe/H] $< -3$ and the trends are due to the
difference of the lifetime of progenitors with different
masses. Homogeneous mixing is required because previous SN yields that
have been used [\eg Woosley \& Weaver (1995, hereafter \cite{woo95});
\citealt{nom97}; \cite{ume02}; and Chieffi \& Limongi (2002, hereafter
\cite{chi02})] show a large scatter in [$\alpha$/Fe] (where $\alpha$
represents $\alpha$-elements, for example, O, Ne, Mg, Si, \eg
\citealt{arg02}). 

However, this interpretation may not be consistent with the Galactic
chemical evolution models that suggest inhomogeneous mixing in such
early phases (\eg \citealt{arg00,tum05}). Also, $r$-process nuclei
observed in the EMP stars show too large scatters
\citep{mcw98,bur00,nor01,hon04} to be reproduced by the
homogeneous mixing model \citep{ish04}, unless there exist a major site
of $r$-process synthesis other than SN explosions (see
\citealt{arg04}, who concluded core-collapse SNe are more preferable
sites of $r$-process elements than neutron-star mergers). 

In the regime of inhomogeneous mixing, \cite{ume05} have succeeded to
reproduce the observed trends of the ratios, [Cr/Fe], [Mn/Fe], [Co/Fe],
and [Zn/Fe], as a result of chemical enrichment of various SN models
including hyper-energetic explosions ($E_{51}=E/10^{51}{\rm ergs}\geq10$:
hypernovae, hereafter HNe). In their approach, variation of $E$ and the
mixing-fallback process are important \citep{ume02,ume05}. The
mixing-fallback model can solve the disagreement between
[$\alpha$/Fe] and [(Fe-peak element)/Fe] (\eg
\citealt{woo95,nak99,chi02,lim03}).

Traditionally, core-collapse SNe were considered to explode with
$E_{51}\sim1$ as SN~1987A \citep{bli00}, SN~1993J \citep{nom93}, and SN~1994I \citep{nom94}
before the discoveries of HNe SN~1997ef (\citealt{iwa00,maz00}) and SN~1998bw
(Patat er al. 2001; \citealt{nak01}). After these discoveries, the number of Pop I HNe
has been increasing, and the association with gamma-ray bursts (GRBs)
has been established as
GRB~980425/SN~1998bw (Galama et al. 1998; Iwamoto et al. 1998; Woosley
et al. 1999; Nakamura et al. 2001a), GRB~030329/SN~2003dh
(Stanek et al. 2003; Hjorth et al. 2003; Matheson et al. 2003; Mazzali
et al. 2003; Lipkin et al. 2004; Deng et al. 2005), and GRB~031203/SN~2003lw
(Thomsen et al. 2004; Gal-Yam et al. 2004; Cobb et al. 2004; Malesani et
al. 2004; Mazzali et al. 2006a). Though it is an interesting issue
how much fraction of the core-collapse SNe explode as HNe (\eg
Podsiadlowski et al. 2004),
non-negligible number of HNe occurred at least at present days.

Nucleosynthesis in HNe is characterized by the large amount of
\Nifs\ production ($M({\rm ^{56}Ni})\gsim0.1$, \eg \citealt{nak01}). Two sites of
\Nifs\ synthesis have been suggested: the shocked stellar core
(\eg \citealt{nak01,mae03}) and the accretion disk surrounding the
central black hole (\eg MacFadyen \& Woosley 1999). We investigate the
former site because the light curve and spectra of SN~1998bw favor
the \Nifs\ synthesis in the shocked stellar core
\citep{mae06a,mae06b,mae06c}.

In this paper, we construct core-collapse SNe models of the Pop III 13
-- 50 $\Msun$ stars for various explosion energies of 
$E=1-40\times 10^{51}$ ergs. By applying the
mixing-fallback model to the HN models, we show that the yields of these
Pop III SNe and HNe are in good agreements with the observed abundance
patterns and trends with reasonably small dispersions \citep{cay04,hon04}. We do not
consider pair-instability SNe because previous studies (\citealt{ume02}; Umeda \&
Nomoto 2003, hereafter \citealt{ume03}; \citealt{ume05}) found that
there has been no evidence of the pair-instability SN abundance patterns in the EMP
stars, although they might have existed before our galaxy become
metal-rich-enough to form low-mass stars (see also \citealt{ohk05} for
core-collapse very massive stars).

In \S~\ref{sec:model}, we describe our progenitor and explosion
models. In \S~\ref{sec:Individual}, we show that the abundance patterns of
the EMP stars are reproduced by HN models and that the abundance patterns of
the VMP stars are reproduced by normal SN models or an IMF-integrated
model. In \S~\ref{sec:trend}, we show that the trends with small
dispersions can be reproduced by Pop III SN models with various
progenitors' masses and explosion energies assuming a SN-induce star
formation. The scatters in our models are almost consistent with the
observed ones. In \S~\ref{sec:improve}, we examine how the agreements of
Sc/Fe, Ti/Fe, Mn/Fe, and Co/Fe are improved by modifying the
distributions of the neutron-proton (n/p) ratio and the density in the
presupernova models. In \S~\ref{sec:discuss}, summary and discussion are
given.

\section{EXPLOSION MODELS}
\label{sec:model}

The calculation method and other assumptions are the same as described
in Umeda \etal\ (2000, hereafter \cite{ume00}), 
\cite{ume02}, and \cite{ume05}. The isotopes in the reaction network for
explosive nuclear burning include 280 species up to $^{79}$Br as in
\cite{ume05}. We calculate the evolutions of Pop III progenitors, whose
main-sequence masses are $\Mms=$ 13, 15, 18, 20, 25, 30, 40, 50 $\Msun$,
and their SN explosions as summarized in Table~\ref{tab:models}. The
mass loss rate from stars with metallicity $Z$ is assumed to be
proportional to $Z^{0.5}$ \citep{kud00}, so that Pop III $Z=0$ models
experience no mass loss. SN hydrodynamical calculations
include nuclear energy generation with the $\alpha$-network. The
yields are obtained by detailed nucleosynthesis calculations
as a post-processing.

\subsection{Energy Injection}

There are various ways to simulate the explosion. For example,
\cite{chi02} assumed an approximate analytic formula to describe the
radiation-dominated shock, although their recent modeling applied full
hydrodynamics (\citealt{lim03}). \cite{woo95} and \cite{lim03} injected energy as a
piston. Their piston model could mimic the time delay until the
deposited energy by neutrino reaches a critical value. However, since
the explosion mechanism of core-collapse SNe have not been well
uncovered, one still does not know what is the most realistic way to
inject explosion energy. Further, the yields do not strongly depend on
how to generate the shock \citep{auf91}. Therefore, in our calculation,
the explosion is initiated as a thermal bomb with an arbitrary explosion
energy, \ie we elevate temperatures of an inner most region of the
progenitor. 

\subsection{Normal Core-Collapse Supernovae and Hypernovae}

\begin{figure*}
\epsscale{1.75}
\plotone{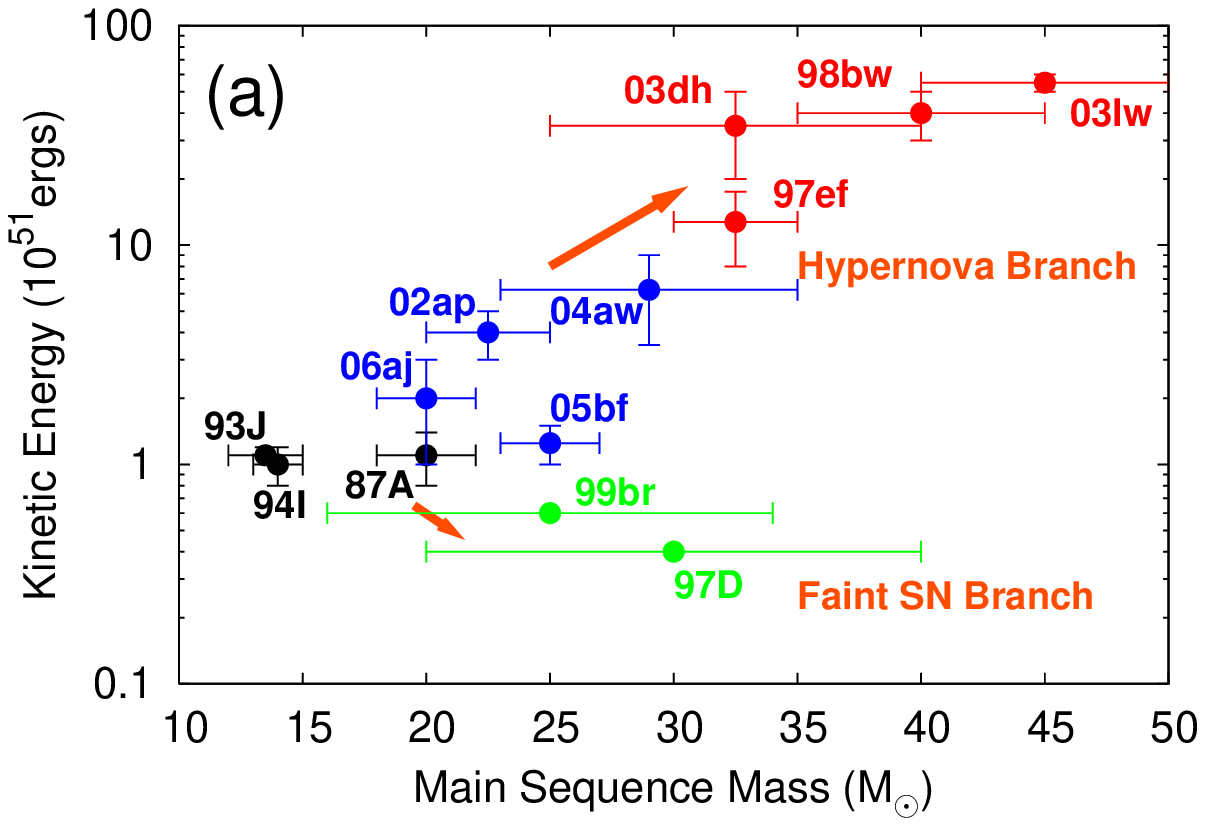}
\plotone{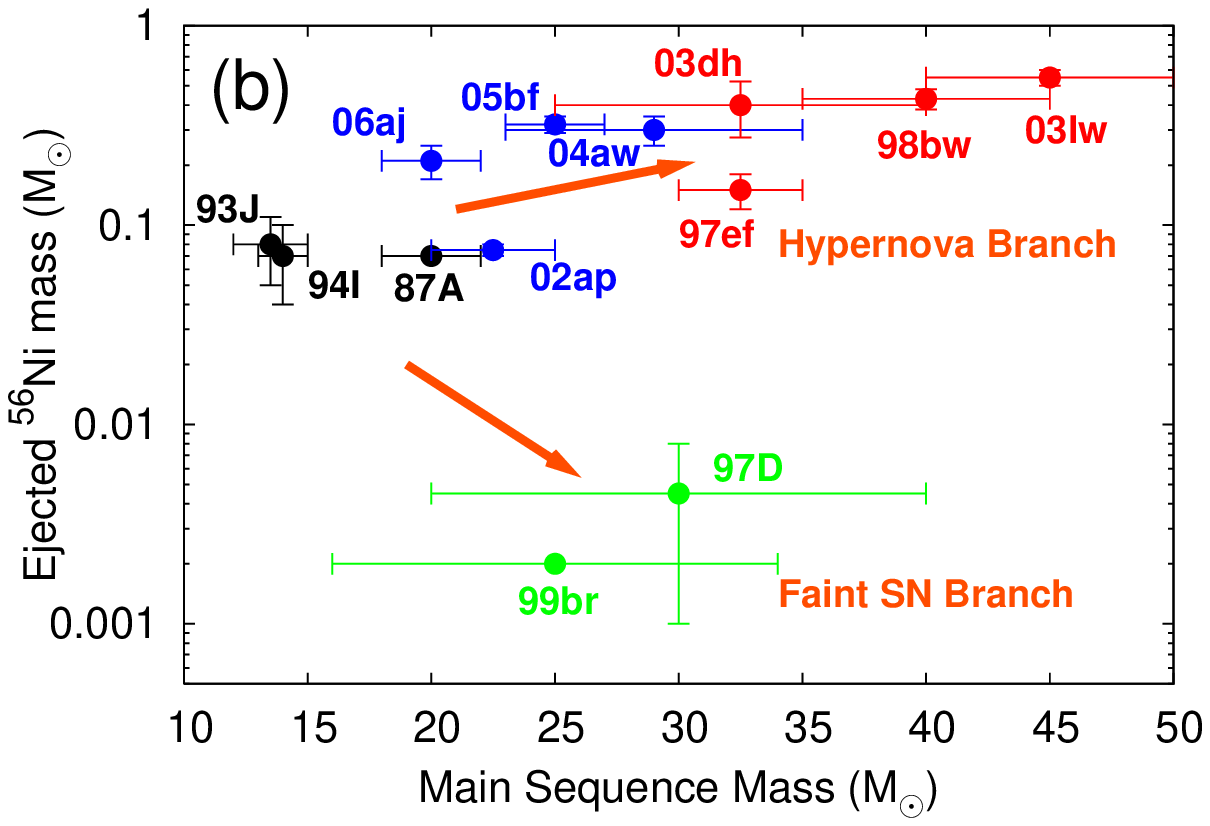}
\figcaption{(a) The explosion energy and (b) the ejected \Nifs\ mass as a
 function of the main sequence mass of the progenitor for core-collapse SNe
 (SN~1987A, \citealt{shi90}, \citealt{bli00}; SN~1993J, \citealt{nom93},
 \citealt{shi94}; SN~1994I, \citealt{nom94}, \citealt{iwa94},
 \citealt{sau06}; SN~1997D, \citealt{tur98}, \citealt{zam03}; SN~1997ef,
 \citealt{iwa00}, \citealt{maz00}; SN~1998bw, \citealt{iwa98},
 \citealt{nak01}; SN~1999br, \citealt{zam03}; SN~2002ap,
 \citealt{maz02}, \citealt{tomita06}; SN~2003dh, \citealt{maz03},
 \citealt{den05}; SN~2003lw, \citealt{maz06}; SN~2004aw,
 \citealt{tau06}; SN~2005bf, \citealt{tom05}; SN~2006aj,
 \citealt{maz06b}).
\label{fig:CCSN}}
\end{figure*}

We determine the explosion energy with referring to the
relation between the main-sequence mass and the explosion energy
($\Mms-E$) as obtained from observations and models of SNe
(Fig.~\ref{fig:CCSN}a). This relation is obtained from Pop I SNe, but
we assume that the same $M_{\rm MS}-E$ relation holds for
Pop III SNe because the Fe core masses are roughly the same between Pop
I and Pop III stars \citep{ume00}. According to the observed relation,
the massive stars with $\Mms \geq 20 \Msun$ are assumed to explode as
hypernovae (HNe), which are hyper-energetic explosions and expected to
leave black holes behind \citep{nom03}, and the explosion energies of
the models for $\Mms=$ 20, 25, 30, 40, 50 $\Msun$ are $E_{51}=$ 10, 10,
20, 30, 40, respectively. The stars smaller than 20 $\Msun$ are
considered to explode as normal SNe with $E_{51}=1$. The explosion
energy of normal SN models ($E_{51}=1$) is consistent with that of
SN~1987A ($E_{51}=1.1 \pm 0.3$, \citealt{bli00}). 

\begin{figure*}
\epsscale{1.3}
\plotone{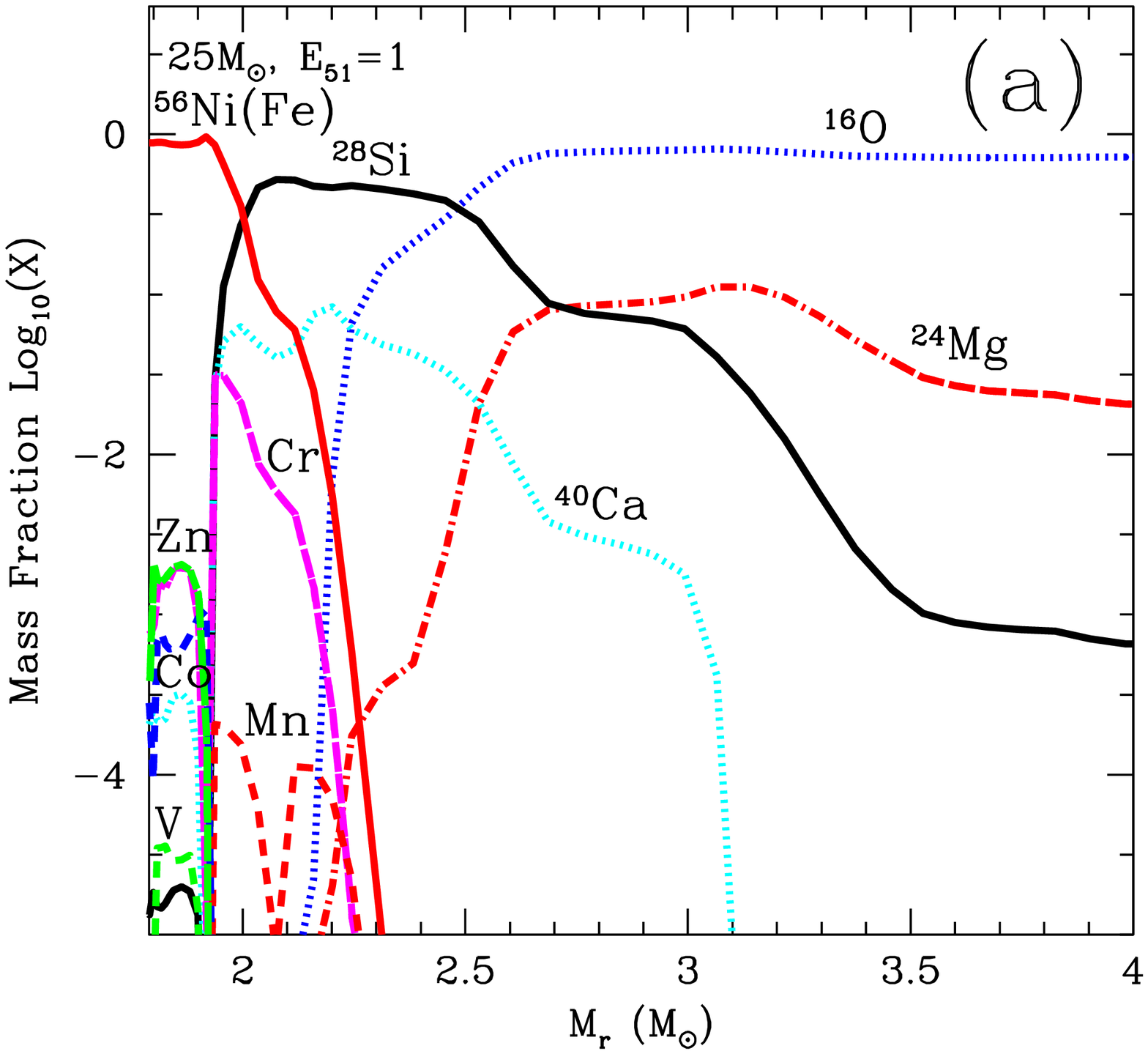}
\plotone{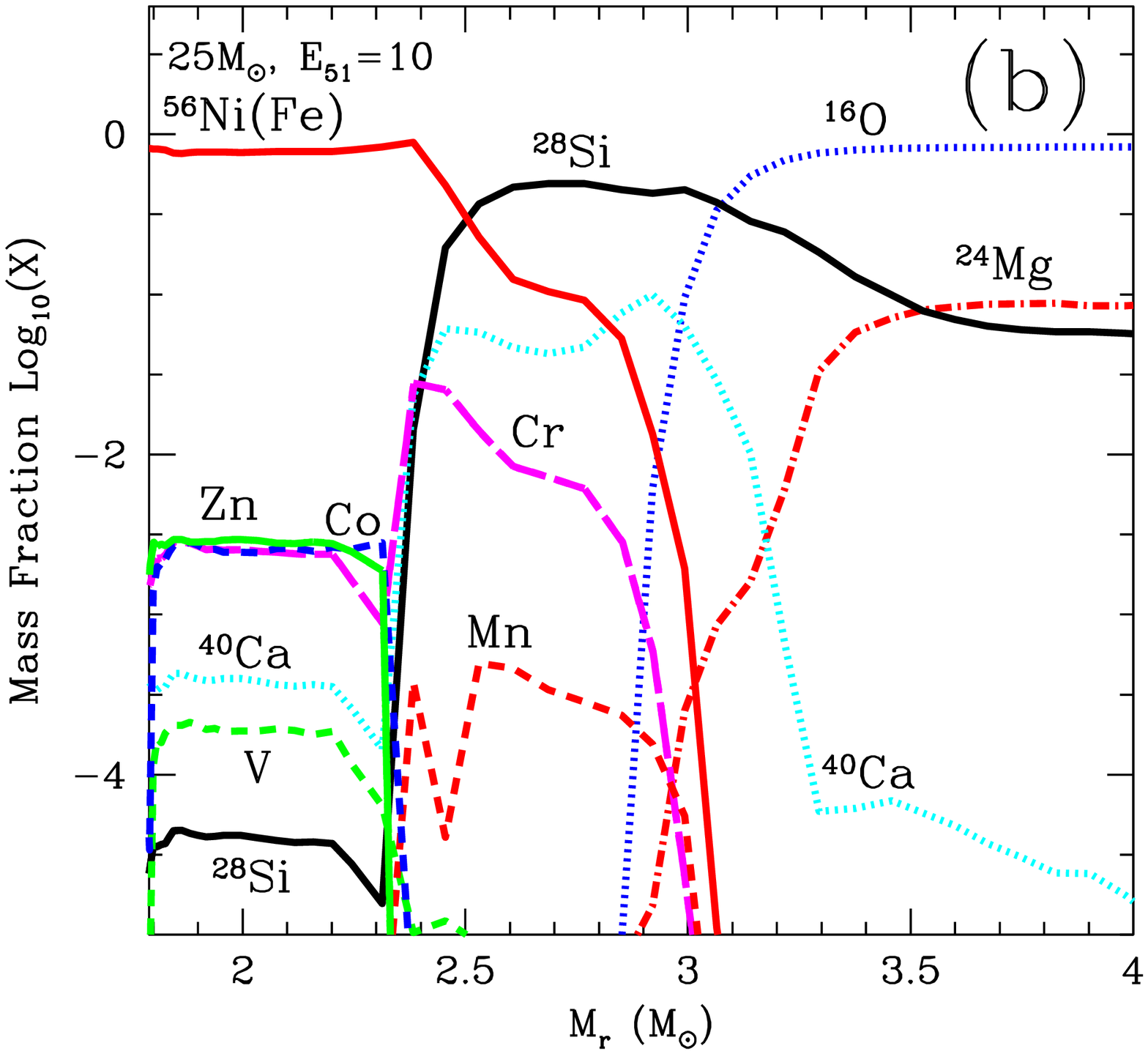}
\figcaption{Abundance distribution against the enclosed mass
$M_r$ after the explosion of Pop III 25 $\Msun$ stars with (a) $E_{51} = 1$
 and (b) $E_{51} = 10$.
\label{fig:SNHN}}
\end{figure*}

Figures~\ref{fig:SNHN}a and \ref{fig:SNHN}b show the abundance
distributions in the
ejecta of the $25\Msun$ normal SN ($E_{51} = 1$) and HN
($E_{51} = 10$) models, respectively. Nucleosynthesis in HNe with large explosion
energies takes place under high entropies and show
the following characteristics \citep{nak01b,nom01,nom04,nom06,ume02,ume05}. 

(1) Both complete and incomplete Si-burning regions shift outward in
mass compared with normal supernovae. As seen in
Figures~\ref{fig:SNHN}a and \ref{fig:SNHN}b, the mass in the complete Si-burning region becomes
larger, while the incomplete Si-burning region does not change much.  As a
result, higher energy explosions produce larger [(Zn, Co,
V)/Fe] and smaller [(Mn, Cr)/Fe].

(2) In the complete Si-burning region of hypernovae, elements produced
by $\alpha$-rich freezeout are enhanced because of high entropies.
Hence, elements synthesized through the $\alpha$-capture process, such
as $^{44}$Ti, $^{48}$Cr, and $^{64}$Ge are more abundant. These species
decay into $^{44}$Ca, $^{48}$Ti, and $^{64}$Zn, respectively.

(3) Oxygen burning takes place in more extended regions for the larger
explosion energy.  Then more O, C, Al are burned to produce a larger amount of
burning products such as Si, S, and Ar. As a result, hypernova
nucleosynthesis is characterized by large abundance ratios of
[(Si, S)/O] (\citealt{ume02b}).

\subsection{Mass Cut}

The mass cut is defined to be a boundary between the
central remnant and the SN ejecta and thus corresponds to the mass of
the compact star remnant. The SN yield is an integration over the ejecta
outside the mass cut. In 1D spherically symmetric models, the mass cut can
be determined hydrodynamically as a function of the explosion energy
\citep{woo95}. However, such 1D hydrodynamical determinations may not be
relevant because SN explosions are found to be aspherical (\eg \citealt{kaw02,leo02,wan02,wan03,mae02,mae06a,mae06b,maz05,chu05}).

We take into account approximately the aspherical effects with a
mixing-fallback model (see Appendix for detail) parameterising the
aspherical SN explosions with three parameters, \ie the initial mass cut
$M_{\rm cut}{\rm (ini)}$, the outer boundary of the mixing region
$M_{\rm mix}{\rm (out)}$ and the ejection factor $f$. The initial
mass of the central remnant is represented by $M_{\rm cut}{\rm (ini)}$. 
During the explosion, an inversion of the abundance distribution and a
fallback of the materials above $M_{\rm cut}{\rm (ini)}$ onto the
central remnant might occur in the aspherical explosions (\eg
\citealt{mae03,nag06,tom06}),
in contrast to the spherical explosions. The inversion is represented by
the mixing of the materials between $M_{\rm cut}{\rm (ini)}$ and 
$M_{\rm mix}{\rm (out)}$ and the fraction of materials ejected from the
mixing region is parameterized by the ejection factor $f$. As a
consequence, the final mass of the central remnant 
$M_{\rm cut}{\rm (fin)}$ is derived with Equation (7) in Appendix.

The parameters of the mixing-fallback model should essentially be
derived from the mechanisms of SN explosions, \eg the rotation, the
asphericity, and the way to inject energy from the central region. However,
the explosion mechanisms of SNe have not been clarified. Thus the
parameters are constrained from the observed Fe (\Nifs) mass or from
the comparison between the yield and the observed abundance pattern.

For normal SN models, we determine the mass cuts to yield 
$M(^{56}{\rm Ni})=0.07\Msun$ because the ejected \Nifs\ mass of the
observed normal SNe are clustered around this value
(Fig.~\ref{fig:CCSN}b, \eg SN~1987A, \citealt{shi90}; SN~1993J,
\citealt{shi94}; SN~1994I, \citealt{nom94}). We also assume that the whole
materials above the mass cut are ejected without the mixing-fallback
process. This corresponds to the ejection factor $f=1$ in the
mixing-fallback model (see Appendix). 

For HN models, we apply the mixing-fallback model
and determine parameters to yield [O/Fe] $=0.5$ (case A) for all HN
models or [Mg/Fe] $=0.2$ (case B) for massive HN models
($\Mms\geq30\Msun$) as described in Appendix. While case A is similar to
\cite{ume05}, case B is to reproduce [Mg/Fe] $\sim 0.2$ plateau for
[Fe/H] $< -3.5$ in \cite{cay04}. The plateau had not been observed
by the previous studies. Since case B has larger amount of fallback, \ie
larger $M_{\rm mix}{\rm (out)}$ and smaller $f$ than case A, both of
small [Mg/Fe] and [Fe/H] is realized in case B. The applied
mixing-fallback parameters are summarised in Tables~\ref{tab:models},
\ref{tab:modelsYe}, and \ref{tab:modelsYelow} and the
final yields are summarised in Tables~\ref{tab:yield}, \ref{tab:yield2},
\ref{tab:yieldYe}, \ref{tab:yield2Ye}, \ref{tab:yieldYelow}, and
\ref{tab:yield2Yelow}.

\subsection{SN-induced star formation}

We assume that [Fe/H] of a star formed by the SN shock compression is
determined by the ratio between the ejected Fe mass $M$(Fe) and the
swept-up H mass $M$(H) \citep{cio88,rya96,tho98,shi98}. According to
\cite{tho98}, the swept-up H mass is given as 
\begin{equation}
 M{\rm (H)}=X{\rm (H)} \cdot M_{\rm SW} = 3.93 \times 10^4 E_{51}^{6/7} n^{-0.24} \Msun
\end{equation}
where $X{\rm (H)}$ is the mass fraction of H in the primordial gases,
$M_{\rm SW}$ is the total amount of the swept-up materials, and $n$ is
the number density of the circumstellar medium. Here we adopt
$X{\rm (H)}=0.752$ as obtained from {\it Wilkinson Microwave Anisotropy
Probe} \citep{spe03} and standard big bang nucleosynthesis \citep{coc04}.
[Fe/H] of the star is approximated as 
\begin{eqnarray}
\nonumber {\rm [Fe/H]} &=& \log_{10}(M{\rm (Fe)}/M{\rm (H)}) \\
\nonumber              & & -\log_{10}{(X{\rm (Fe)}/X{\rm (H)})_\odot}  \\
&\simeq& \log_{10}\left({M({\rm Fe})/{\Msun}\over{E_{51}^{6/7}}}\right)-C. \label{eq:SNSF}
\end{eqnarray}
Here $(X{\rm (Fe)}/X{\rm (H)})_\odot$ is the solar abundance ratio in mass
between Fe and H \citep{and89} and $C$ is assumed to be a
constant value of 1.4, which corresponds to $n\sim 75$ $ {\rm cm^{-3}}$.
Resulting values of [Fe/H] are summarised in Tables~\ref{tab:models},
\ref{tab:modelsYe}, and \ref{tab:modelsYelow}. [Fe/H] of stars formed
from the ejecta of HNe and normal SNe can be consistent with the
observed [Fe/H] of the EMP and VMP stars, respectively.

\section{COMPARISONS WITH INDIVIDUAL STARS}
\label{sec:Individual}

\subsection{EMP stars}
\label{sec:EMP}

\cite{cay04} provided abundance patterns of 35 metal-poor stars with
small error bars for $-4\lsim{\rm [Fe/H]}\lsim-2$. Each EMP
star may be formed from the ejecta of a single Pop III SN, although some
of them might be the second or later generation stars. The yields of SNe
with the progenitors of [Fe/H] $<-3$ can be well-approximated by
those of Pop III SNe since they are similar \citep{woo95,ume00}. In this
subsection, the theoretical yields are compared with the averaged abundance
pattern of four EMP stars, CS~22189-009, CD-38:245, CS~22172-002 and
CS~22885-096, which have low metallicities ($-4.2<{\rm [Fe/H]}<-3.5$) and
normal [C/Fe] $\sim 0$. Stars with large [C/Fe] ($\sim +1$), called
C-rich EMP stars, are discussed in \cite{ume03}, \cite{ume05}, and
\citealt{tom06}. The origin of some of those stars may be a faint SN,
being different from those of [C/Fe] $\sim 0$ stars (see \S~\ref{sec:C}). 
Also some of the C-rich EMP stars might be originated from the mass
transfer from the C-rich companion in close binaries (\eg
\citealt{aok02,bee05,rya05,coh06}).

\begin{figure*}
\epsscale{2.1}
\plotone{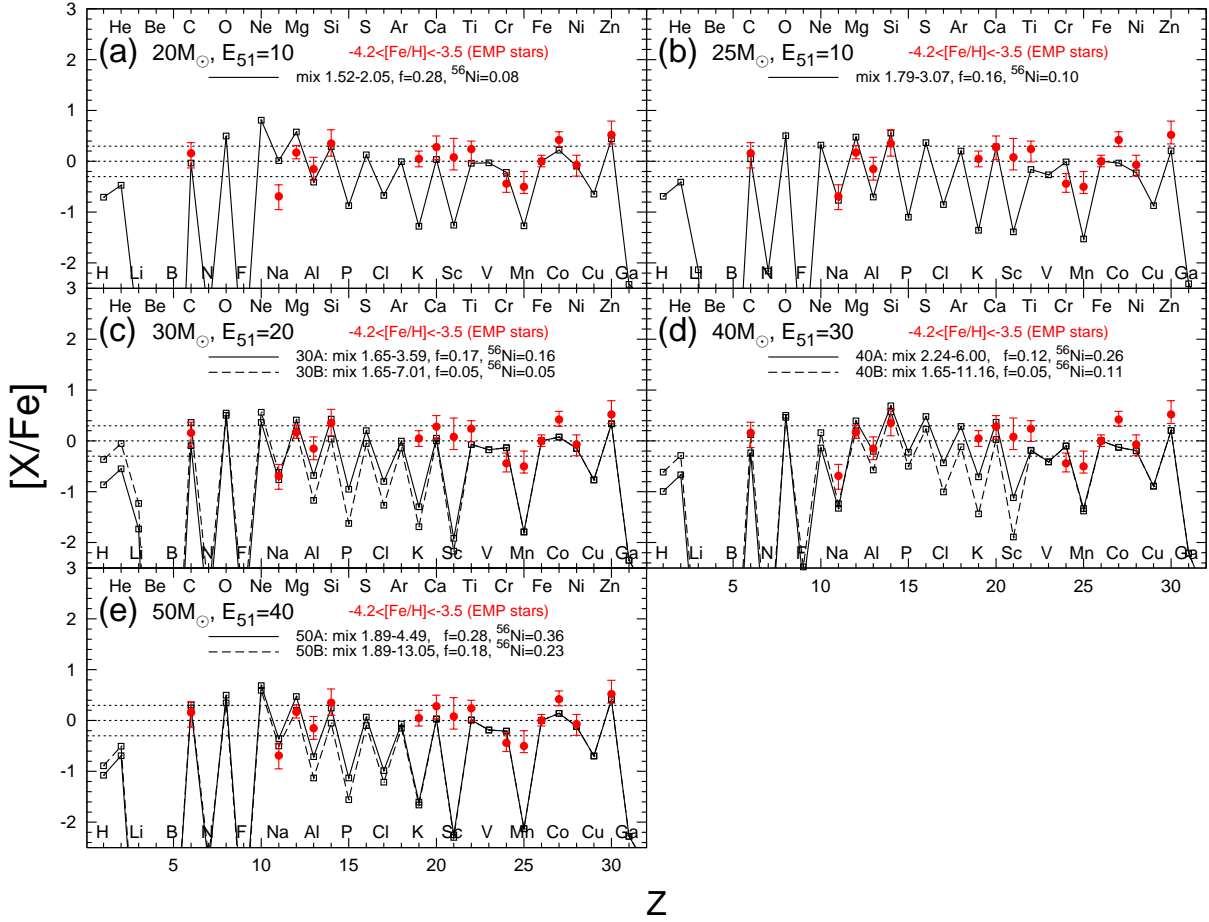}
\figcaption{The comparison between the abundance pattern of the EMP
stars given by \cite{cay04} ({\it filled circles with error bars})
and the theoretical individual HN yields (case A: {\it solid line}, case B:{\it
dashed line}) with the mixing-fallback model. The parameters are shown
in the figures and in Table~\ref{tab:models}. Figures show the
comparisons with
(a) $M_{\rm MS}=20\Msun$, $E_{51}=10$,
(b) $M_{\rm MS}=25\Msun$, $E_{51}=10$,
(c) $M_{\rm MS}=30\Msun$, $E_{51}=20$, 
(d) $M_{\rm MS}=40\Msun$, $E_{51}=30$, and 
(e) $M_{\rm MS}=50\Msun$, $E_{51}=40$.
\label{fig:EMP}}
\end{figure*}

Comparisons between the HN yields and the abundance pattern of the EMP stars
are made in Figures~\ref{fig:EMP}a-\ref{fig:EMP}e. In
the mixing-fallback model, both [(Fe-peak elements)/Fe] and [$\alpha$/Fe]
give good agreements with the observations, except for some elements, \eg
K, Sc, Ti, Cr, Mn, and Co. Possible ways to improve these elements are
discussed in \S~\ref{sec:improve}.

\subsection{VMP Stars}
\label{sec:VMP}

\cite{cay04} also provided the abundance patterns of the VMP stars
whose metallicities ([Fe/H] $\sim -2.5$) are higher than the EMP stars. 
The observed abundance pattern is represented by the averaged abundance
pattern of five stars BD+17:3248, HD~2796, HD~186478, CS~22966-057 and
CS~22896-154, which have relatively high metallicities 
($-2.7<{\rm [Fe/H]}<-2.0$). Since these metallicities correspond
to those of normal SN models according to Equation (\ref{eq:SNSF}) for the
SN-induced star formation model, we first compare the observations with
normal SN yields (Figs.~\ref{fig:VMP}abc). The mass cuts of normal SN
models are determined so that the ejected Fe (\Nifs) mass is 
$M({\rm Fe})=0.07\Msun$.

On the other hand, most VMP stars are considered to have the
abundance pattern averaged over IMF and metallicity of the progenitors,
thus we also compare with the IMF-integrated yield (Fig.~\ref{fig:IMF}). 
Since the yields of SNe with the progenitors of [Fe/H] $<-3$ are
quite similar to those of Pop III SNe \citep{woo95,ume00}, we use the
Pop III yields for these stars as well. The IMF integration
is performed from 10 $\Msun$ to 50 $\Msun$ with 8 models, 13, 15, 18,
20, 25, 30, 40, 50 $\Msun$ and the extrapolations. We make the power-law
IMF integrations as follows:
\begin{equation}
 \phi(M)=KM^{-(1+a)}
\end{equation}
where $\phi(M)dM$ is the number of stars within the mass range of
[$M,M+dM$], $K$ is a normalization constant, and $a$ is an integration
index (Salpeter IMF has $a=1.35$). The integration is performed and
normalized by the total amount of gases forming stars as follows:
\begin{equation}
 X({\rm A})={\int^{50\Msun}_{0.07\Msun} X_M({\rm A}) M_{\rm ej}(M) \phi(M) dM
  \over{\int^{50\Msun}_{0.07\Msun} M \phi(M) dM}}
\label{eq:integration}
\end{equation}
where $X({\rm A})$ is an integrated mass fraction of an element, A,
$X_M({\rm A})$ is mass fraction of A in a model interpolated between
the nearest models, and $M_{\rm ej}(M)$ is an ejected mass interpolated
between the nearest models or the nearest model and the edge of the 
IMF-integrated mass range.  Here we assume $M\leq10M_\odot$ and
$M=50M_\odot$ stars do not yield any materials as type II SNe or HNe,
i.e., $M_{\rm ej}(M\leq10M_\odot)=M_{\rm ej}(50M_\odot)=0M_\odot$.

Comparing the integrated yield with the Salpeter's IMF with the abundance
pattern of the VMP stars (Fig.~\ref{fig:IMF}), most elements show
reasonable agreements, except for N, K, Sc, Ti, and Mn. The integrated
yields are summarised in Table~\ref{tab:IMF}.

\subsubsection{Nitrogen \& Oxygen}
\label{sec:NO}

\begin{figure*}
\epsscale{1.55}
\plotone{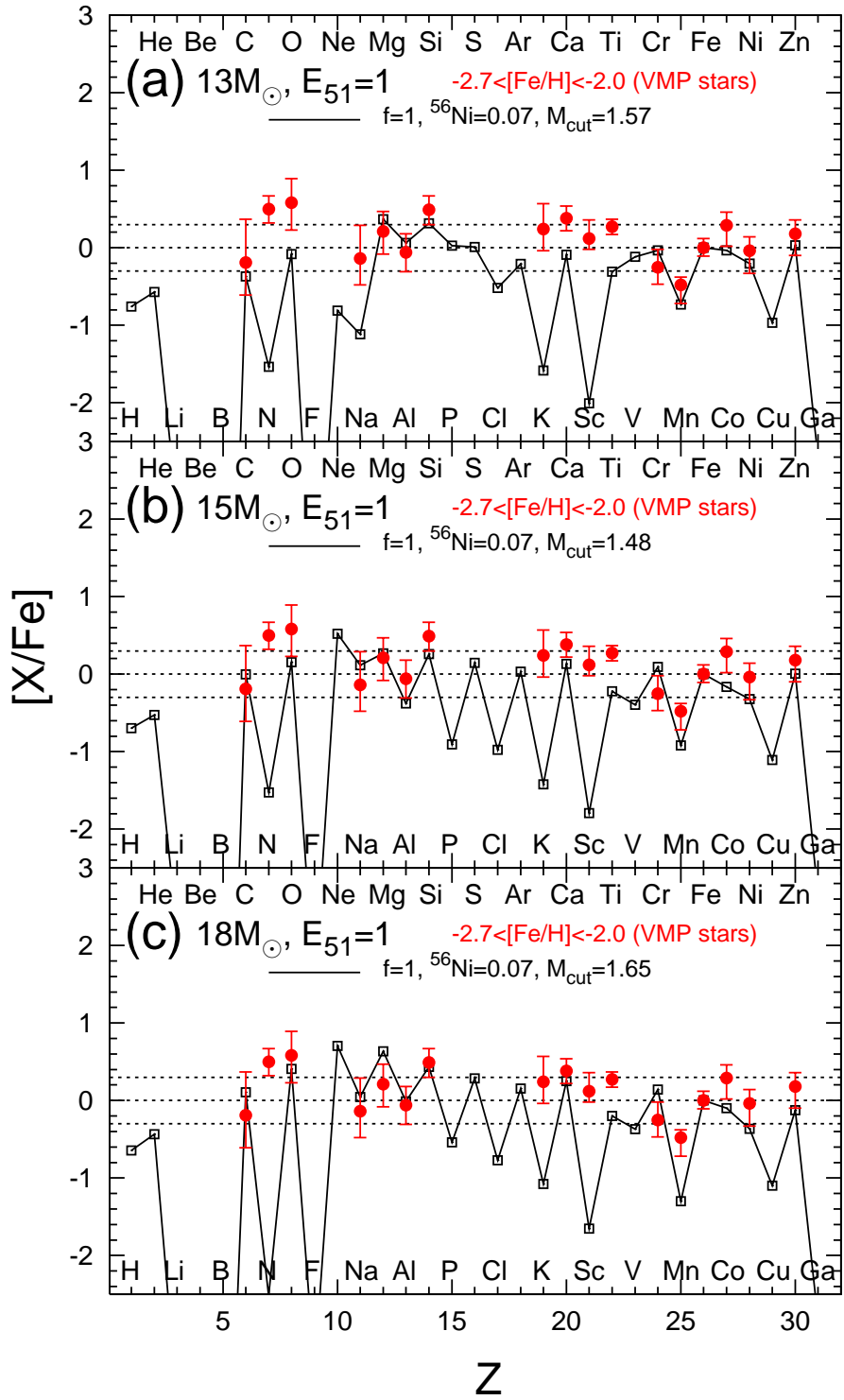}
\figcaption{The comparison between the
abundance pattern of the VMP stars given by \cite{cay04} ({\it filled
circles with error bars}) and the theoretical individual normal SN
yields ({\it solid line}).
Figures show the comparisons with
(a) $M_{\rm MS}=13\Msun$, $E_{51}=1$, 
(b) $M_{\rm MS}=15\Msun$, $E_{51}=1$, and
(c) $M_{\rm MS}=18\Msun$, $E_{51}=1$.
\label{fig:VMP}}
\end{figure*}

\begin{figure*}
\epsscale{2.1}
\plotone{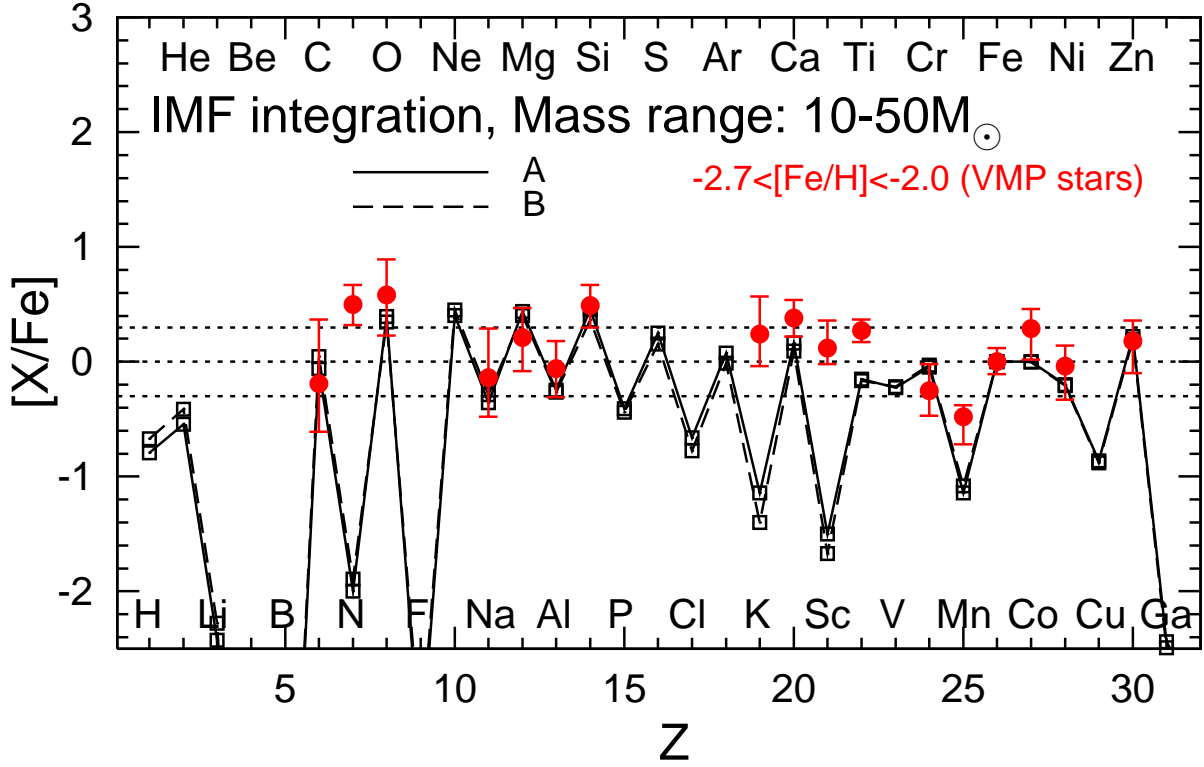} 
\figcaption{Same as Figure~\ref{fig:VMP}, but for the IMF-integrated
yield of Pop III SNe from 10$\Msun$ to 50 $\Msun$. The mixing-fallback
model is applied for HN models, not for normal SN models. In case A
({\it solid line}), all HN models are determined their mixing-fallback
parameters so that [O/Fe] $=0.5$, but in case B ({\it dashed line}),
massive HN models larger than $\Mms=30\Msun$ has mixing-fallback
parameters so that [Mg/Fe] $=0.2$.
\label{fig:IMF}}
\end{figure*}

Figures~\ref{fig:VMP}a-\ref{fig:VMP}c and \ref{fig:IMF} show that N is
underproduced in our models. 
There are two possible explanations (1) and (2) for this discrepancy:

(1) N was actually underproduced in the Pop III SN as in our models, but
       was enhanced as observed during the first dredge-up in the
       low-mass red-giant EMP stars (\eg
       \citealt{wei04,sud04}). Observationally, the EMP and VMP stars 
       in \cite{cay04} are giants and some of them show evidences of the
       deep mixing, \ie the dilution of Li and the low $^{12}$C/$^{13}$C ratio
       \citep{spi06}. On the other hand, the EMP and VMP stars with
       no evidence of the deep mixing show relatively small [N/Fe] ($\sim-1$,
       \citealt{spi05}). However, [N/Fe] in our models are even smaller
       than the smallest [N/Fe] observed in the EMP and VMP stars 
       \citep{spi05}. Thus the following mechanism might be important.

(2) N was enhanced in massive progenitor stars before the SN
       explosion. N is mainly synthesized by the mixing between the
       He convective shell and the H-rich envelope (\eg
       \citealt{ume00,iwa05}). The mixing can be enhanced by
       rotation \citep{lan92,heg00,mae00}. Suppose that the Pop III
       SN progenitors were rotating faster than more metal-rich stars
       because of smaller mass loss, then [N/Fe] was enhanced as
       observed in the EMP stars.

[O/Fe] of the 18 $\Msun$ and the IMF-integrated model are in good agreement
with the observations (Figs.~\ref{fig:VMP}c and \ref{fig:IMF}), while
[O/Fe] of the 13 and 15 $\Msun$ models are lower than the observations 
(Figs.~\ref{fig:VMP}a and \ref{fig:VMP}b). In the abundance determination of
O, however, uncertain hydrodynamical (3D) effects are important
\citep{nis02} and \cite{cay04} applied the 3D correction for dwarfs to
the metal-deficient giants. If the observed values of [O/Fe] in the figures
are correct, this may indicate that the contribution of a single normal
SN from a small-mass progenitor to the chemical enrichment in the VMP
stars is small. 

We assumed all massive stars with $\Mms\geq 20\Msun$ explode as HNe.
However, there is a suggestion that the fraction of HNe to whole SNe 
($\epsilon_{\rm HN}$) is $\epsilon_{\rm HN} \sim0.5$
(\citealt{kob05}). If $\epsilon_{\rm HN} \sim0.5$, [(C, N, O)/Fe] and
[Zn/Fe] are slightly larger and smaller, respectively, than the case
with $\epsilon_{\rm HN} \sim1$, 
but these are still in good agreement with the observation (Fig.~12 in
\citealt{nom06}). On the other hand, if the contribution of the faint SNe 
($\Mms\geq 20\Msun$) to reproduce the abundance patterns of the C-rich
EMP stars is large enough (see \S~\ref{sec:C}; \citealt{ume03,ume05,tom06}), 
[(C, N, O)/Fe] is enhanced. This is because the faint SNe produce large
[(C, N, O)/Fe] due to a small amount of Fe ejection. The
contribution of the faint SNe, however, might be small, since [Mg/Fe] is
close to the upper limit of the observations without the contribution of
the faint SNe. In order to estimate the ratios of HNe and faint SNe relative
to all core-collapse SNe, further investigations are necessary.

\section{TRENDS WITH METALLICITY}
\label{sec:trend}

In \S~\ref{sec:EMP} and \ref{sec:VMP}, we show that the observed
abundance patterns can be reasonably reproduced by the mixing-fallback
model. \cite{cay04} showed not only the abundance patterns of individual
stars but also the existence of certain trends of the abundance ratios
with respect to [Fe/H]. In this section, we compare the observed trends with
our models in Tables~\ref{tab:models}-\ref{tab:yield2}.

\begin{figure*}
\epsscale{2.1}
 \plotone{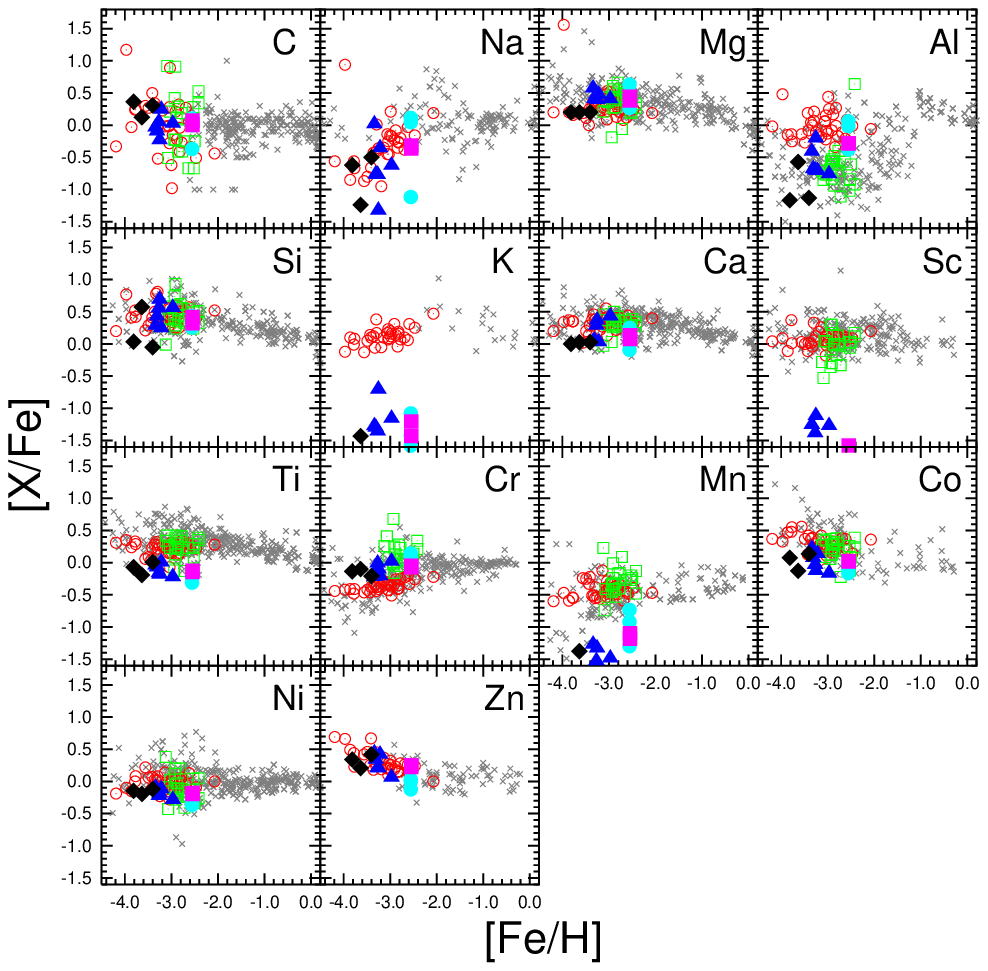}
\figcaption{The comparison between the [X/Fe] trends of observed stars [the
previous studies (\eg \citealt{gra91,sne91,edv93,mcw95a,mcw95b,rya96,mcw97,car00,pri00,gra03,ben03}); {\it cross}, \cite{cay04}; {\it open circle},
\cite{hon04}; {\it open square}] and those of individual stars models
(normal SNe: {\it filled circle}; HNe with case A: {\it filled
triangle}, HNe with case B: {\it filled rhombus}) and IMF integration ({\it
filled square}). The parameters are shown in Table~\ref{tab:models}.
\label{fig:trend}}
\end{figure*}

In Figure~\ref{fig:trend}, the observed abundance ratios [X/Fe] against
[Fe/H] are compared with yields of individual SN models in
Table~\ref{tab:models} and the IMF-integrated yield described in
\S~\ref{sec:VMP}. Here [Fe/H] of individual SN models are determined by
Equation~(\ref{eq:SNSF}), while [Fe/H] of the IMF-integrated abundance
ratios are assumed to be same as normal SN models ([Fe/H] $\sim -2.6$).
We note that the observed abundance ratios of most elements are roughly
constant for $-2.5 \lsim$ [Fe/H] $< -1$. This can be interpreted
that the SN ejecta had been mixed homogeneously in the halo at $-2.5 \lsim$
[Fe/H]. This is consistent with the chemical evolution models in
\cite{ish99}, \cite{arg00}, \cite{tum05}, \cite{nom06}, and \cite{kob05}. 

\begin{figure*}
\epsscale{2.1}
\plotone{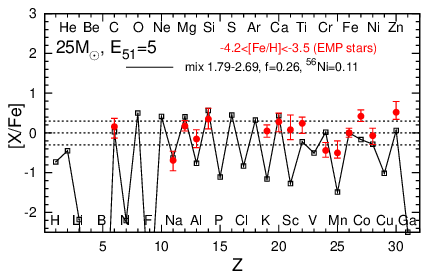}\figcaption{Same as Fig.~\ref{fig:EMP}, but for 
$M_{\rm MS}=25\Msun$, $E_{51}=5$.
\label{fig:25E5}}
\end{figure*}

According to the SN-induced star formation model (Eq.~\ref{eq:SNSF}),
our models cluster around [Fe/H] $\sim -3.5$ and only a few model exists
around [Fe/H] $\sim -3$, because we applied only one explosion energy
for each mass. In reality, the explosion energies of HNe may depend,
\eg on the rotation of the progenitors, even if the progenitors' masses of
HNe are similar. The progenitors of SNe~1997ef and 2003dh, for instance,
have similar masses, but the explosion energies are $1\times 10^{52}$ ergs 
and $4\times 10^{52}$ ergs, respectively \citep{iwa00,maz03,den05}. 
Variations of the explosion energy for the same stellar mass lead
to variations of [Fe/H]. In order to produce a model with 
${\rm [Fe/H]}\sim-3$, therefore, we add the $25\Msun$ model with
$E_{51}=5$ (see Fig.~\ref{fig:25E5} for its abundance pattern).

The trends of most elements can be well
reproduced by our models, except for K, Sc, Ti, Cr, Mn, and Co. In
the followings, we discuss the trend of each element in more detail.  

\subsection{Carbon}
\label{sec:C}

\begin{figure*}
\epsscale{2.1}
\plotone{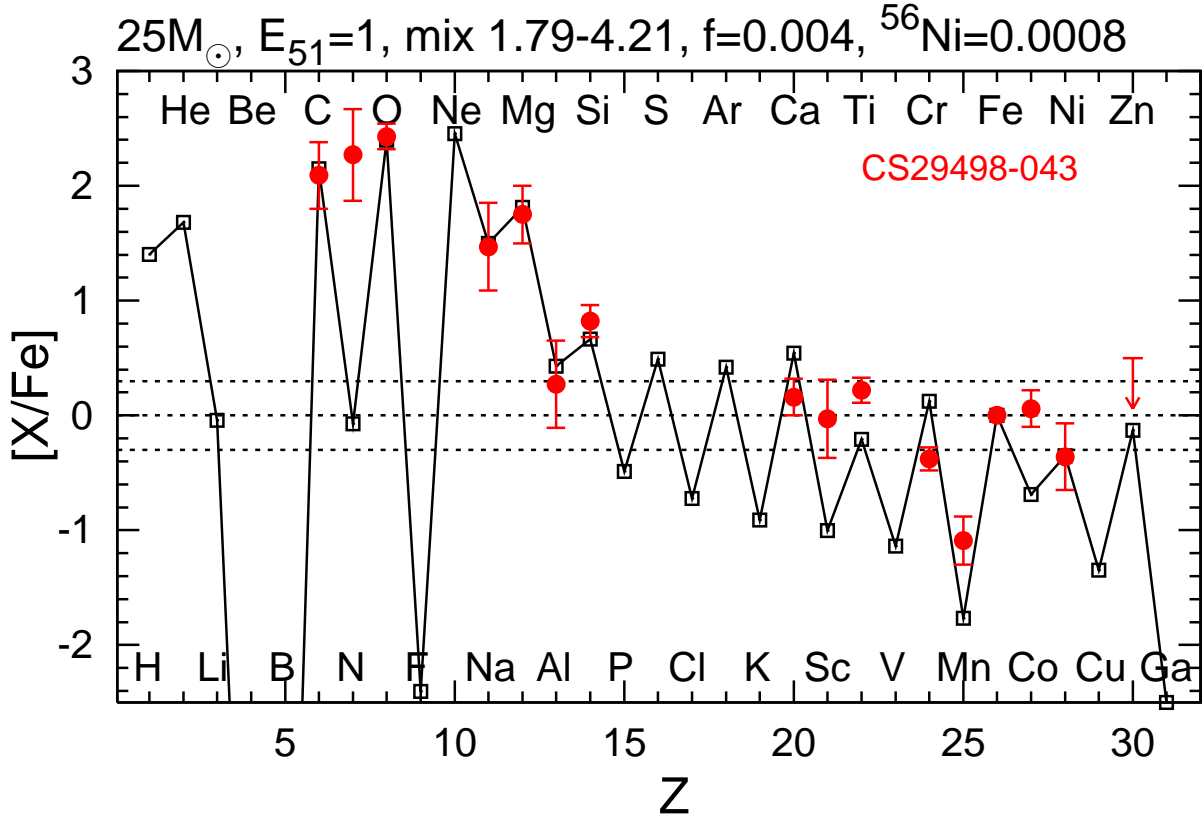}\figcaption{The comparison between the
abundance pattern of the C-rich EMP star (CS~29498-043: \citealt{aok04},
{\it filled circles with error bars}) and the theoretical faint SN
yields (25F: {\it solid line}).
The mixing-fallback parameters are determined so as to reproduce the
abundance pattern of CS~29498-043. \label{fig:faint}}
\end{figure*}

The ratio C/Fe in our models are clustered around [C/Fe] $\sim 0$,
while the observed [C/Fe] of the EMP stars are scattered (see
Fig.~\ref{fig:trend}). The large [C/Fe] ($\gsim 0.5$) can be interpreted
as originated from the faint SNe that are characterised
by a small ejection factor $f$ and the resultant large [(C, N, O)/Fe] \citep{ume03,ume05}. In fact,
Figure~\ref{fig:faint} shows that the abundance pattern of the C-rich EMP stars
(CS~29498--043: \citealt{aok04}) can be reproduced by our 25 $\Msun$
faint SN model with a normal explosion energy $E_{51}=1$ and small
$f=0.004$ (model 25F in Tables~\ref{tab:models}-\ref{tab:yield2}). 
In this faint SN model, N/Fe is too small but can be enhanced as
described in \S~\ref{sec:NO}. 

We should note that the large [Co/Fe] of CS~29498--043 is difficult to
be reproduced by the faint SN model of $E_{51}=1$. The parent SN of
CS~29498--043 might be a faint SN with high-entropy jets \citep{ume05,tom06}. 

Alternative explanation is that the production of C in other sites is
important, that is, C is
produced not only from SNe but also from mass-losing Wolf-Rayet (WR)
stars. Fast rotating stars may undergo strong mass-loss and enter the WR
phase, even if their metallicities are considerably low \citep{mey05}.
Also, the C-enhancement can be realized by transferring mass from the
C-rich binary companion (\eg \citealt{aok02,rya05,coh06}).

\subsection{The Even-Z Elements: Mg, Si, Ca, and Ti}
\label{sec:even}

\subsubsection{Magnesium, Silicon \& Calcium}
\label{sec:MgSiCa}

The ratio Mg/Fe in \cite{cay04} is less-scattered around [Mg/Fe]
$\sim0.2$ which is smaller than [Mg/Fe] $\sim0.5$ obtained by \cite{hon04} and other previous studies. Case A models,
whose ejection factor $f$ is set to produce [O/Fe] $=0.5$, appear around
$-3.5 <$ [Fe/H] $< -2.6$ and $0.1 <$ [Mg/Fe] $< 0.6$. 
These are larger by 0.1 dex than observed [Mg/Fe] in \cite{cay04}
but consistent with these in the previous studies. We also consider case
B models to reproduce the [Mg/Fe] $\sim0.2$ plateau in \cite{cay04} at
[Fe/H] $<-3.5$, although the plateau had not been observed by the
previous studies.

[(Si, Ca)/Fe] in \cite{cay04} and \cite{hon04} are slightly
less-scattered than in previous studies. [(Si, Ca)/Fe] in our models are
in good agreements with the observations and the widths of the scatter
in our models are consistent with the observations. 

The small scatters of [(Mg, Si, Ca)/Fe] in our present model are
different from previous yields (\citealt{nom88, woo95}; Thielemann et
al. 1996). This
difference stems from the following different assumption. The previous
yields assumed the ejected Fe mass ($M({\rm Fe})\sim0.07\Msun$) depends
weakly on the progenitors' masses. Therefore [$\alpha$/Fe] in massive
star models depends mainly on the $\alpha$-element abundances, which
strongly depend on the stellar mass. This leads to the large
scatters. In the present stars, the EMP stars with [Fe/H] $< -3$ are
produced by HNe only ($\Mms\geq20\Msun$), and those HN models produce Fe
much more than $\sim 0.07\Msun$. Instead, we assume [O/Fe] $=0.5$ or
[Mg/Fe] $=0.2$ for $\Mms\geq20\Msun$ models. Our approach suggests that
the observed small scatter of [$\alpha$/Fe] implies that larger amount
of Fe is produced in more massive stars. This is consistent with the
observations that typical HNe eject larger amount of Fe than normal SNe
(Fig.~\ref{fig:CCSN}b).

\subsubsection{Titanium}
\label{set:Ti}

[Ti/Fe] in our models is smaller than the observations. There
are no clear trend in [Ti/Fe] in our models as in the observations and
the scatter is similar to the observations. [Ti/Fe] may be enhanced by
nucleosynthesis in high-entropy environments as will be discussed in
\S~\ref{sec:improve} (a ``low-density'' modification, see also
\cite{ume05}) or in a jet-like explosion \citep{nag03,mae03}.

\subsection{The Odd-Z Elements: Na, Al, K, and Sc}
\label{sec:odd}

\subsubsection{Sodium \& Aluminum}
\label{sec:NaAl}

The abundances of Na and Al in \cite{cay04} are different from previous
studies. This is partly because \cite{cay04} took into account non-LTE
(NLTE) effects ($\Delta X({\rm Na}) = -0.5$ dex and 
$\Delta X({\rm Al}) = +0.65$ dex). 

The \cite{cay04} result shows a trend in [Na/Fe] against [Fe/H]. In
Figure~\ref{fig:trend}, our results also show such a trend, although Na/Fe
is slightly smaller than the observations in the HN models, especially
in $\Mms=40\Msun$ models. Because all our models are Pop III models, the
trend is not due to the metallicity effect but due to the combination of the
progenitors' masses and the explosion energies. The trend stems from the
fact that more massive stars produce smaller $X$(Na) and 
that more energetic explosions burn more Na.

[Al/Fe] is also smaller by 0.5 dex than in \cite{cay04}, especially in
the HN model. However our models are in good agreements with [Al/Fe] in
\cite{hon04}. The difference in the observed [Al/Fe] between
\cite{cay04} and \cite{hon04} may be mainly because \cite{hon04} does not
take into account NLTE effects. The larger $M_{\rm mix}{\rm (out)}$ in
case B enhances Na/Fe but reduces Al/Fe. This is because Na and Al
are mainly synthesized near the outer and the inner edges of the O-rich
region, respectively. Only [Al/Fe] of model 40B is in reasonable
agreement with \cite{cay04} because the 40 $\Msun$ progenitor star
produce large $X$(Al).

Both Na and Al are mostly synthesized in C shell-burning, and the
produced amount depends on the overshooting at the edge of the
convective C-burning shell \citep{iwa05}. In the present presupernova
evolution models, no overshooting is included. If the NLTE correction of
Al is correct, it might be needed to consider convective overshooting in
the C-burning shell. This could enhance Na/Mg and Al/Mg, thus
weakening the odd-even effect in the abundance patterns and leading to a
better agreement with \cite{cay04}. 

\subsubsection{Potassium \& Scandium}
\label{sec:KSc}

[K/Fe] and [Sc/Fe] in our models are much smaller than in \cite{cay04} and
\cite{hon04}. Possible improvements are discussed in detail in
\S~\ref{sec:improve}. K/Fe is slightly enhanced by the ``low-density''
modification as described in \S~\ref{sec:improve}, but still not large
enough. Iwamoto et al. (in preparation) suggests that the model with large
$Y_{\rm e}$ ($> 0.5$) in the inner region can produce enough K.

Sc/Fe can be enhanced by the ``low-density'' modification (see
\S~\ref{sec:improve}; \cite{ume05}). Further enhancement can be
realized if $Y_{\rm e}>0.5$. Recently, \cite{pru04,pru05} and \cite{fro04}
calculated nucleosynthesis based on the core-collapse SN simulations
with neutrino transport. In their models, neutrino absorption enhances
$Y_{\rm e}$ near the mass cut of the ejecta. According to their results,
Sc/Fe in the normal SN model is enhanced to [Sc/Fe] $\sim 0$ due to the
large $Y_{\rm e}$ ($\gsim0.5$) and high-entropy ($s/k_{\rm B}\gsim20$,
where $s$ denotes the the entropy per nucleon and $k_{\rm B}$ is the
Boltzmann constant).

\subsection{The Iron-Peak Elements: Cr, Mn, Co, Ni, Zn}
\label{sec:Fe-peak}

\subsubsection{Chromium, Manganese, \& Cobalt}
\label{sec:CrMnCo}

[Cr/Fe] in our models is larger than \cite{cay04} but consistent
with \cite{hon04}. The difference between \cite{cay04} and \cite{hon04}
stems from the use of the different Cr lines (\ion{Cr}{1}: \cite{cay04}
and \ion{Cr}{2}: \cite{hon04}) in obtaining [Cr/Fe]. It is still
uncertain which line is better to use in estimating [Cr/Fe]. If
\ion{Cr}{1} gives reliable abundance, Cr in our models is overproduced,
although the trend in our models is similar to the observations. Since
Cr is mostly produced in the incomplete Si-burning region, the 
size of this region relative to the complete Si-burning region should be
smaller than the present model in order to produce smaller Cr/Fe. We
also need to examine the nuclear reaction rates related to the synthesis
of ${\rm ^{52}Fe}$ that decays into ${\rm ^{52}Cr}$.

[Mn/Fe] and [Co/Fe] in our models are smaller than the observations,
although the trend of [Co/Fe] in our models is similar to the
observations. The negligibly small dependence of [Mn/Fe] on [Fe/H] in
\cite{cay04} is different from previous observations that [Mn/Fe] significantly
decreases toward smaller [Fe/H] \citep{mcw95a,mcw95b}. They can be
improved by the n/p modification as discussed in \S~\ref{sec:Ye}. 
Also, Mn can be efficiently enhanced by a neutrino process (see
\S~\ref{sec:Yelowtrend}; \citealt{woo95}; T.~Yoshida \etal, in preparation). Therefore the Mn/Fe ratio is important to
constrain the physical processes during the explosion.

\subsubsection{Nickel \& Zinc}
\label{sec:NiZn}

[Ni/Fe] and [Zn/Fe] in our models are in good agreement with the
observations, although [Ni/Fe] is slightly smaller than the
observation. Ni/Fe is higher if $M_{\rm {cut}}{\rm (ini)}$ is smaller (\ie the
mass cut is deeper) because $^{58}$Ni, a main isotope of stable Ni, is mainly
synthesized in a deep region with $Y_{\rm e} < 0.5$. However, a smaller 
$M_{\rm {cut}}{\rm (ini)}$ tends to suppress Zn/Fe (see Appendix), thus
requiring more energetic explosions to produce large enough Zn/Fe.

The good agreement of [Zn/Fe] with observations strongly supports the
SN-induced star formation model and suggests that the EMP stars with smaller
[Fe/H] are made from the ejecta of HNe with higher explosion energies
and larger progenitor's masses. This is because higher explosion
energies lead to larger [Zn/Fe] and smaller [Fe/H]. 

Other possible production sites of Zn include the neutrino-driven wind
from a proto-neutron star (\eg \citealt{hof96,pru05,fro04,wan06a,wan06b}) 
and the accretion disk of a black hole (\eg \citealt{pru04,pru04b,fro06}). Further studies are needed to see how
large the contributions of these sites to the Zn production are. In
order to reproduce the trend and small scatter of [Zn/Fe], there must be
some ``hidden'' relations between the explosion energy and
nucleosynthesis in the neutrino-driven wind or the accretion disk models.

If the VMP stars with [Fe/H] $\sim -2.5$ is made
from normal SNe with $E_{\rm 51}=1$, Zn is underproduced in
our models. Nucleosynthesis studies with neutrino transport
\citep{hof96,pru04,pru05,fro04} suggested that Zn in the normal SN model is
enhanced to [Zn/Fe] $\sim 0$.
However, the enhancement is not large enough to explain the large [Zn/Fe]
($\gsim 0.5$) in the EMP stars. 

\section{IMPROVED MODELS FOR Sc, Ti, Mn, \& Co}
\label{sec:improve}

In this section, we present the models based on the modified
presupernova distributions of the ``n/p ratio (\ie $Y_{\rm e}$)'' and
the ``density''. We then discuss how these modifications improve Sc,
Ti, Mn, and Co.
The IMF-integrated yields, the parameters, and the yields of the individual
SNe are summarised in Tables~\ref{tab:IMF}-\ref{tab:yield2Yelow}.

\subsection{Neutron-Proton Ratio}
\label{sec:Ye}

In the above discussion, we assume that $Y_{\rm e}$ in the presupernova model
is kept almost constant during the explosion. However, recent studies (\eg
\citealt{ram00,lie05,fro04,bur06a,bur06b}) have suggested that $Y_{\rm e}$ may be significantly
varied by the neutrino process during explosion. Further, the region, where the
neutrino absorption and $Y_{\rm e}$ variation occur, is Rayleigh-Taylor
unstable because of neutrino heating, so that there exists a large
uncertainty in $Y_{\rm e}$ and its distribution. 

We apply the following $Y_{\rm e}$ profile that was found to produce
reasonable results in \cite{ume05}, \ie $Y_{\rm e}=0.5001$ in the
complete Si burning region and $Y_{\rm e}=0.4997$ in the incomplete Si
burning region. The $Y_{\rm e}$ profile is modified by adjusting the isotope ratios
of Si. According to the recent explosion calculations with the neutrino
effect (\eg \citealt{ram00,bur06b}), materials with large 
$Y_{\rm e}$ ($\sim 0.54$) may be ejected. However, $Y_{\rm e}$ might be
diluted by mixing, and our $Y_{\rm e}$ profile might mimic such
dilution. The adopted high $Y_{\rm e}$ in the complete Si-burning
region and low $Y_{\rm e}$ in the incomplete Si-burning
region lead to large Co/Fe and Mn/Fe, respectively.

\begin{figure*}
\epsscale{2.1}
\plotone{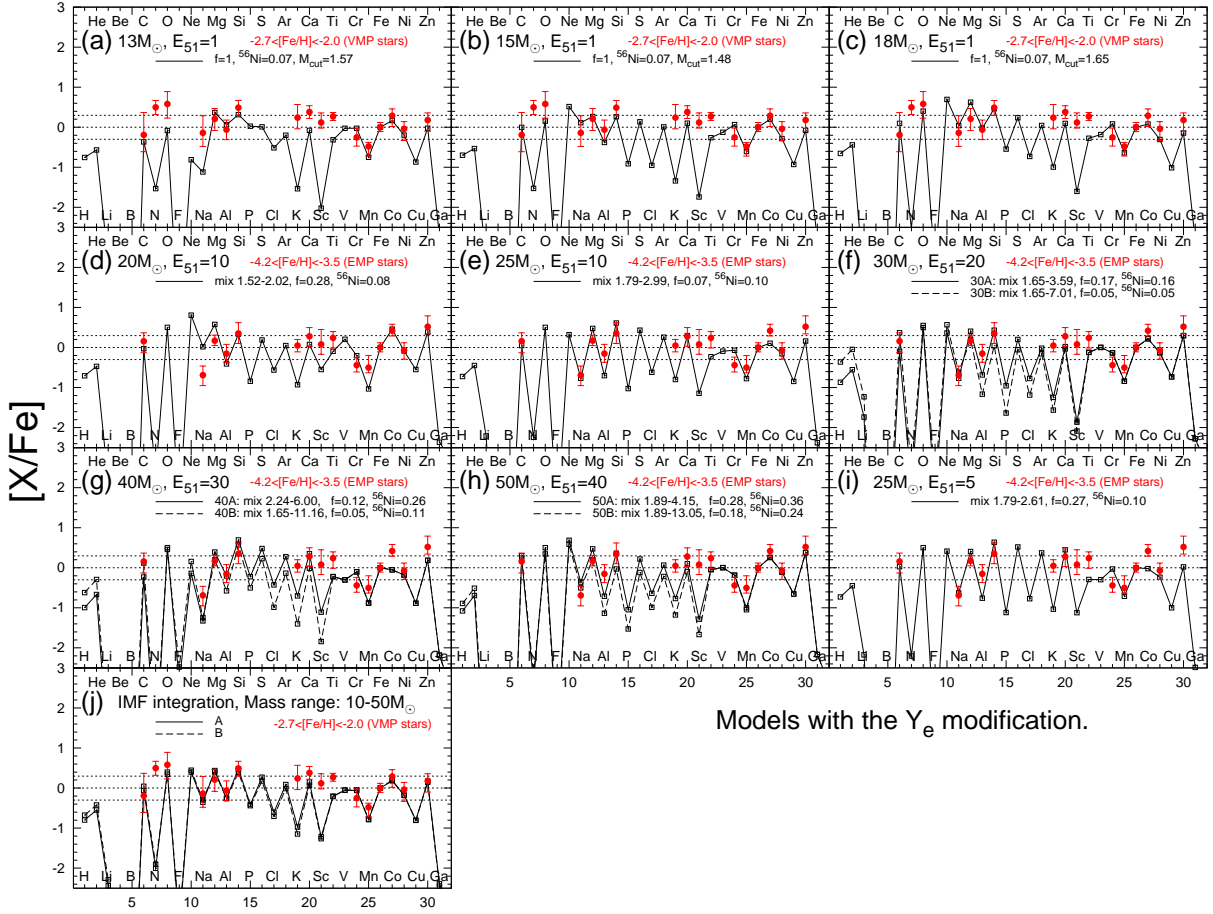}
\figcaption{Same as Fig.~\ref{fig:EMP}, but models are applied the
 $Y_{\rm e}$ modification. Figures (a)-(c) show the comparisons between
the abundance pattern of the VMP stars ({\it filled circles with error
bars}) and the normal SN models with the $Y_{\rm e}$ correction ({\it solid line}),
(a) $M_{\rm MS}=13\Msun$, $E_{51}=1$,
(b) $M_{\rm MS}=15\Msun$, $E_{51}=1$, and
(c) $M_{\rm MS}=18\Msun$, $E_{51}=1$.
Figures (d)-(i) show the comparisons between the abundance pattern of
the EMP stars ({\it filled circles with error bars}) and the HN models with
the $Y_{\rm e}$ modification (case A: {\it solid line}, case B:{\it
dashed line}),
(d) $M_{\rm MS}=20\Msun$, $E_{51}=10$,
(e) $M_{\rm MS}=25\Msun$, $E_{51}=10$,
(f) $M_{\rm MS}=30\Msun$, $E_{51}=20$, 
(g) $M_{\rm MS}=40\Msun$, $E_{51}=30$,
(h) $M_{\rm MS}=50\Msun$, $E_{51}=40$, and
(i) $M_{\rm MS}=25\Msun$, $E_{51}=5$.
Figures (j) shows the comparisons between
the abundance pattern of the VMP stars ({\it filled circles with error
bars}) and the IMF-integrated yield of Pop III SNe from 10$\Msun$ to 50
$\Msun$ with (a)-(h) models.
\label{fig:Ye}}
\end{figure*}

Figures~\ref{fig:Ye}a-\ref{fig:Ye}j show better agreements between the
models and the observations. Here we applied the same 
$M_{\rm cut}{\rm (ini)}$ as in the models without the $Y_{\rm e}$
modification. The IMF-integrated yield, the parameters, and the yields
of the individual SNe are summarised in
Tables~\ref{tab:IMF}-\ref{tab:yield2Ye}. Since the variation of  
$Y_{\rm e}$ is considered to be due to the $\nu$-process, the $Y_{\rm e}$
modification might be applied to any core-collapse SNe. 

\subsection{Low Density}
\label{sec:low}

In the ``low-density'' modification, the density of the presupernova
progenitor is artificially reduced without changing the total mass. 
\cite{ume05} assumed that such a low density would be realized if
the explosion is induced by multiple jets consisted of the primary
weak jets and the main strong SNe jets. The primary weak jets
expand the interior of the progenitor before the SN explosions driven by
the main jets (as described in Appendix in \citealt{ume05}). 
Alternate explosion mechanism that realizes ``low-density'' is
a delayed explosion (\eg \citealt{fry06}, who investigated
explosive nucleosynthesis induced by a black hole forming collapse
comparing a direct collapse and a delayed collapse caused by
fallback). The ``low-density'' is presumed to be realized in the
jet-like or delayed explosions involving fallback, thus
being applied only to the HN models but not to the normal SN models. 

Other mechanism to realize ``low-density'' was suggested by
\cite{chi02}, \ie the L model using ``low'' $^{12}$C($\alpha$,
$\gamma$)$^{16}$O rate \citep{cau88}. They showed that the ``low''
$^{12}$C($\alpha$, $\gamma$)$^{16}$O rate leads the higher C abundance
(\ie larger C/O ratio) and more active C-shell burning. As a
consequence, the average density in the complete and incomplete
Si-burning region is lower than the ``high''  $^{12}$C($\alpha$,
$\gamma$)$^{16}$O rate case \citep{cau85}. If this mechanism is valid, the
``low-density'' will be realized in whole SNe. However, the large C/O
ratio leads to overproduce Ne and Na at the solar metallicity
\citep{woo93,nom97,imb01}.

\begin{figure*}
\epsscale{2.1}
\plotone{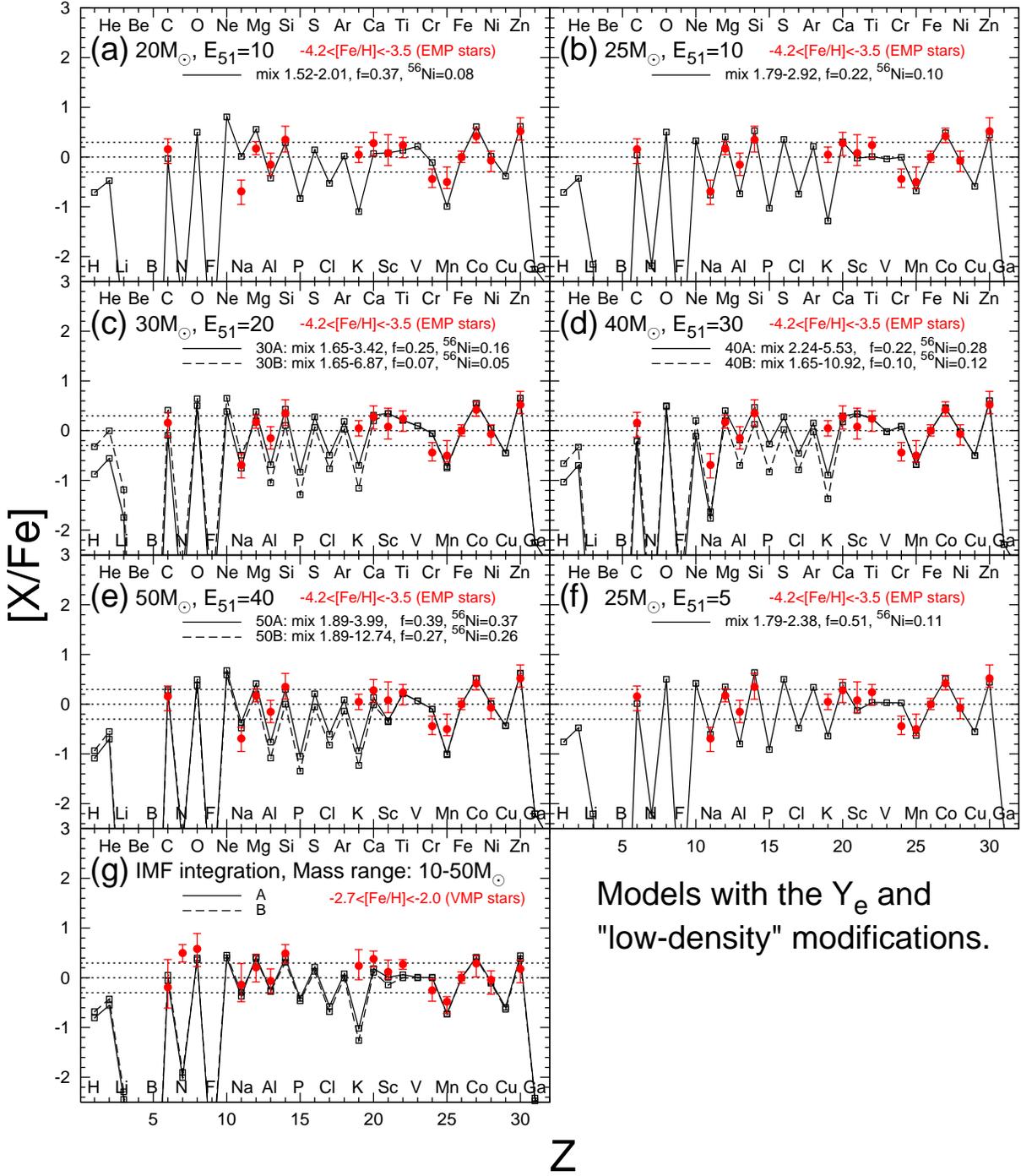}
\figcaption{Same as Fig.~\ref{fig:EMP}, but models are applied the
 $Y_{\rm e}$ and ``low-density'' modifications. 
Figures (a)-(f) show the comparisons between the abundance pattern of
the EMP stars ({\it filled circles with error bars}) and the HN models with the
 $Y_{\rm e}$ and ``low-density'' modifications (case A: {\it solid
line}, case B: {\it dashed line}),
(a) $M_{\rm MS}=20\Msun$, $E_{51}=10$,
(b) $M_{\rm MS}=25\Msun$, $E_{51}=10$,
(c) $M_{\rm MS}=30\Msun$, $E_{51}=20$, 
(d) $M_{\rm MS}=40\Msun$, $E_{51}=30$,
(e) $M_{\rm MS}=50\Msun$, $E_{51}=40$, and
(f) $M_{\rm MS}=25\Msun$, $E_{51}=5$.
Figures (g) shows the comparisons between
the abundance pattern of the VMP stars ({\it filled circles with error
bars}) and the IMF-integrated yield of Pop III SNe from 10$\Msun$ to 50
$\Msun$ with normal SN (Figs.~\ref{fig:Ye}abc) and (a)-(e) HN models. \label{fig:Yelow}}
\end{figure*}

Explosive nucleosynthesis in the model with the ``low-density''
modification takes place at higher entropy ($s/k_{\rm B}\sim30-40$) 
than in the model without modification ($s/k_{\rm B}\sim10-20$), thus
enhancing the $\alpha$-rich freeze-out compared with the standard
model. As a result, the Sc/Fe and Ti/Fe ratios are particularly enhanced. 
Figure~\ref{fig:Yelow}a shows that the Sc/Fe and Ti/Fe ratios in the EMP stars
can be better reproduced by the model of $\Mms=20\Msun$ and $E_{51}=10$ with
the ``low-density'' modification, whose presupernova density is reduced
by a factor of 2. Here we applied the $Y_{\rm e}$ modification and the
same $M_{\rm cut}{\rm (ini)}$ as in the models without the $Y_{\rm e}$
modification. 

The other HN models with both the $Y_{\rm e}$ and
``low-density'' modifications are also in good agreement with the
abundance pattern of the EMP stars (Figs.~\ref{fig:Yelow}b-\ref{fig:Yelow}g).
Because the degree of ``low-density'' is likely to be different in each
explosion, we assume the ``low-density'' factor so that each model
reproduces [Sc/Fe] of the EMP stars as given in Table~\ref{tab:modelsYelow}.
The IMF-integrated yield, the parameters, and the yields of
the individual SNe are summarised in Tables~\ref{tab:IMF} and
\ref{tab:modelsYelow}-\ref{tab:yield2Yelow}. 

The good agreements suggest that the $Y_{\rm e}$ and ``low-density''
modifications might actually
be realized in the SN explosions. In the present study, we consider the global
enhancement of entropy due to the jet-like or delayed explosion,
but such high-entropy region is also realized locally in 
the neutrino-driven wind or the accretion disk. Neutrino-driven wind
nucleosynthesis \citep{pru05} for $Y_{\rm e}=0.5245$ and 
$s/k_{\rm B}=26.9$ produces [Sc/Fe] ($\sim+0.3$) being consistent with
that of the EMP stars but smaller [Co/Fe] and [Zn/Fe] than those of the EMP
stars; for $Y_{\rm e}=0.5$ and $s/k_{\rm B}=18.4$, smaller
[Sc/Fe] ($\sim-1$) is produced than that of the EMP stars. Accretion disk
nucleosynthesis ($s/k_{\rm B}=20$ and $40$, \citealt{pru04})
produces much larger [Sc/Fe] ($\sim+1$) than that of the EMP stars and
does not produce simultaneously [Co/Fe] $\sim0.5$ and [Zn/Fe]
$\sim0.5$ as observed in the EMP stars.

The above models show that it seems difficult to reproduce the
overall abundance pattern of the EMP stars only with nucleosynthesis in the
neutrino-driven wind or the accretion disk. However, if some
contributions from these sites can be added to our models, it could
enhance Sc/Fe.
In order to obtain [Sc/Fe] $\sim0$, the SN ejecta should have 
$M{\rm (Sc)}/M{\rm (Fe)}\sim (X{\rm (Sc)}/X{\rm (Fe)})_\odot\sim3\times10^{-5}$
\citep{and89}. 
When about $0.1\Msun$ of Fe and little Sc are ejected as in our model of
$\Mms=25\Msun$ and $E_{51}=10$, [Sc/Fe] $\sim0$ is obtained by the
ejection of extra $M{\rm(Sc)}\sim3\times10^{-6}\Msun$ from either the
neutrino-driven wind or the accretion disk.

\subsection{Trends}
\label{sec:Yelowtrend}

\begin{figure*}
\epsscale{2.1}
\plotone{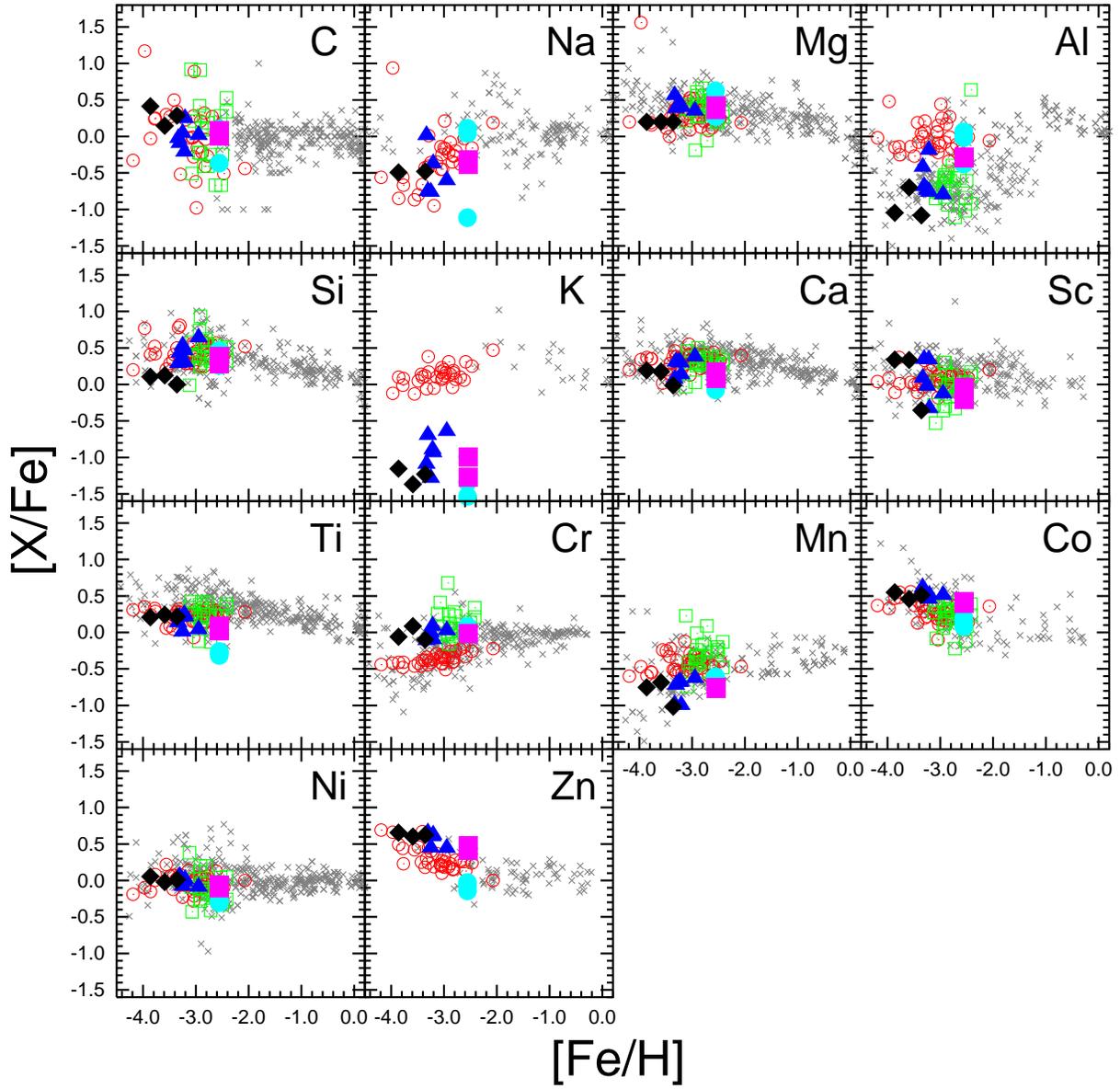}
\figcaption{Same as Fig.~\ref{fig:trend}, but for the comparisons
with models applied the $Y_{\rm e}$ and ``low-density'' modifications. 
\label{fig:trendYelow}}
\end{figure*}

Figure~\ref{fig:trendYelow} shows the trends resulting from the
$Y_{\rm e}$ modification made for all models and the ``low-density''
modification applied only for HN models. Each comparison between the model and the
observation is shown in Figures~\ref{fig:Ye}a-\ref{fig:Ye}c, and
\ref{fig:Yelow}a-\ref{fig:Yelow}g. While [(K, Cr)/Fe]
still do not explain the observations well, the trends of [(Sc, Ti, Mn,
Co)/Fe] are in much better agreement with the observations than the
models without the modifications. Other
elements also show good agreements and small scatters. 

The Sc/Fe and Ti/Fe ratios in normal SN models are lower than the
observations. This is partly because the ``low-density'' modification is not
applied to normal SNe. To reproduce the observed high Sc/Fe and Ti/Fe
ratios, there might be some mechanisms to realize the high entropy
($s/k_{\rm B}\sim30-40$) even for normal SNe. In this connection, it is
interesting to note that the recent X-Ray Flash GRB~060218 was found to
be associated with SN~2006aj (\eg \citealt{pia06}). The properties of
SN~2006aj are close to normal SNe;
the progenitor's main-sequence mass and the explosion energy were
estimated to be $\Mms\sim20\Msun$ and $E_{51}\sim2$ \citep{maz06b}. This suggests that the ``low-density''
modification could be realized in some normal SNe. Such explosions might
contribute to enhance Sc/Fe and Ti/Fe. 

[Cr/Fe] and [Mn/Fe] in our models are larger and smaller than
\cite{cay04}, respectively, while [Co/Fe] is in good agreement with
\cite{cay04}. [Mn/Fe] could also be enhanced by the following
$\nu$-process in the complete Si-burning region and the subsequent
radioactive decay:
\begin{eqnarray}
 {\rm ^{56}Ni} + \nu &\rightarrow&  {\rm ^{55}Co} + \nu' + {\rm p} \\
 {\rm ^{55}Co} &\rightarrow& {\rm ^{55}Mn} + \gamma + {\rm e^+} 
\end{eqnarray}
(\citealt{woo95}; T.~Yoshida \etal, in
preparation). 

\section{SUMMARY \& DISCUSSION}
\label{sec:discuss}

In this paper we performed the hydrodynamical and nucleosynthesis
calculations of Pop III 13 -- 50 $\Msun$ core-collapse SNe and provided
the yields by adopting the mixing-fallback model. 
We show that our yields are consistent with the observed
abundance patterns of the EMP and VMP stars \citep{cay04,hon04}. The
trends of [X/Fe] vs. [Fe/H] with small scatters can be reproduced by our
models as a sequence resulting from the combination of different
progenitors' masses ($\Mms$) and explosion energies ($E$). This is because
we adopt the empirical relation that a larger amount of Fe is ejected by
massive HNe ($\Mms\geq20\Msun$) than normal SNe. This indicates that the
observed trends with small scatters do not necessarily mean the
homogeneous mixing in the interstellar medium, but can be reproduced by the
``inhomogeneous'' chemical evolution model, in which the EMP stars are
enriched by the individual SNe with different ($\Mms$, $E$).

In our model, yields of more massive HNe correspond to the EMP stars with
smaller [Fe/H] ($< -3$). We should stress that this is not because
HNe were dominant among Pop III SNe, but because the second
generation stars produced by Pop III core-collapse SNe with higher
explosion energies tend to have smaller [Fe/H] as 
${\rm [Fe/H]} \simeq \log_{10}\left({M({\rm Fe})\over{\Msun}}/{E_{\rm 51}^{6/7}}\right) -C$ (Eq.~\ref{eq:SNSF}).

Almost all stars in \cite{cay04} and \cite{hon04} can be reasonably well
reproduced by core-collapse SNe yields, but none by the pure pair-instability SN yields
(\eg \citealt{ume02,heg02}). In other words, the EMP stars in \cite{cay04}
show no clear features of top-heavy IMF. Further, the VMP stars ([Fe/H]
$\sim -2.5$) can be well-reproduced by integrating the yields of Pop III
SNe over the Salpeter's IMF. This also implies that the IMF of
Pop III stars was not top-heavy, but
approximately Salpeter's. To constrain the IMF of Pop III stars, we need
more complete set of the observed data as well as inhomogeneous chemical
evolution models which properly take into account our model, especially the
$M_{\rm MS}$-[Fe/H] relation (see, \eg \citealt{arg00,arg02,tum05}).

We also investigate the yields of the models with the $Y_{\rm e}$ and
``low-density'' modifications suggested in \cite{ume05} and
show that the yields are in better agreement with the
observed abundance patterns than those without
such modifications. The good agreements suggest that such modifications
might be realized in the actual core-collapse SN explosions. We suggest
that the $Y_{\rm e}$ and ``low-density'' modifications are actualized by
the neutrino effects and the jet-like or delayed explosions,
respectively. 

The neutrino-driven wind and the accretion disk are other possible
nucleosynthesis sites for Sc because they can actualize the $Y_{\rm e}$
($\gsim0.5$) and
``low-density'' modifications. However, nucleosynthesis in the
neutrino-driven wind and the accretion disk might not be dominant sites
of Fe synthesis because they can not reproduce the abundance ratios
among the Fe-peak elements of the EMP stars. We thus suggest that the
progenitor's presupernova nucleosynthesis and explosive nucleosynthesis are
predominant synthesis sites of most elements, while some elements (\eg
Sc, Ti, Mn, and Co) can be enhanced by the $Y_{\rm e}$ and
``low-density'' modifications.

\acknowledgements

We would like to thank C. Kobayashi for valuable discussion.
The author N.T. is supported through the JSPS (Japan Society
for the Promotion of Science) Research Fellowship for Young Scientists.
This work has been supported in part by the Grant-in-Aid for
Scientific Research (17030005, 17033002, 18104003, 18540231) and the
21st Century COE Program (QUEST) from the JSPS and MEXT of Japan.

\appendix

\section*{APPENDIX: MIXING-FALLBACK MODEL}

The mixing-fallback model proposed by \cite{ume02} and \cite{ume03} can
successfully reproduce the abundance patterns of the hyper metal-poor (HE0107--5240:
\citealt{chr02,ume03}, HE1327--2326: \citealt{fre05,iwa05}) and EMP
\citep{ume02,ume05} stars. The mixing-fallback model assumes the following situation: first, inner materials are mixed by
some mixing process (\eg Rayleigh-Taylor instabilities and/or aspherical
explosions) during the shockwave propagations in the star (\eg
\citealt{hac90,kif03}). Later, some fraction of materials in
the mixing region undergoes fallback onto the central remnant by gravity
(\eg \citealt{woo95,iwa05}), and the rests are ejected into interstellar
space. The fallback mass
depends on the explosion energy, the gravitational potential, and
asphericity. 

The mixing-fallback model can solve a problem associated with the ratios
between the Fe-peak elements and Fe in the EMP stars, [(Fe-peak)/Fe]. The
large Zn/Fe ratio implies an energetic explosion as a HN and a deep mass
cut, but these lead to eject too large amount of Fe to reproduce small
[Fe/H] and the large enough [$\alpha$/Fe] ratio, if ones assume that
the whole material above the mass cut are ejected. However, one can
realize both the deep mass cut and the small amount of ejected Fe with
the mixing-fallback model. 

In the spherical models, \cite{iwa05} found that the fallback takes place if
the explosion energy is less than $E_{51} \sim 1$ for the 25~$\Msun$ star
and obtained a relation between $E_{51}$ and the final remnant mass,
$M_{\rm cut}{\rm (fin)}$, \ie smaller $E_{51}$ leads to a larger 
$M_{\rm cut}{\rm (fin)}$. For example, $E_{51}=0.74$ and $0.71$ for the
25~$\Msun$ star lead to the final remnant mass 
$M_{\rm cut}{\rm (fin)}=5.8\Msun$ and $6.3\Msun$, respectively. 
It is difficult for a spherical HN explosion to initiate a
fallback, although for a larger star the fallback can occur even for a
larger explosion energy because of a deeper gravitational potential. For
instance, the fallback is found to occur for $E_{51} < 2$ in the
50~$\Msun$ star. 

However, fallback can take place not only for relatively low
energy explosions but for very energetic jet-like explosions
\citep{mae03,tom06}. In fact, \cite{tom06} have simulated jet-induced
explosions and showed that the resultant yields can reproduce the
abundance patterns of the EMP stars as the mixing-fallback model. The
mixing-fallback model mimics such aspherical explosions, although the
spherical model tends to require larger explosion energies than the jet
model to obtain similar yields \citep{mae03,tom06}.

\begin{figure*}
\epsscale{2.3}
\plottwo{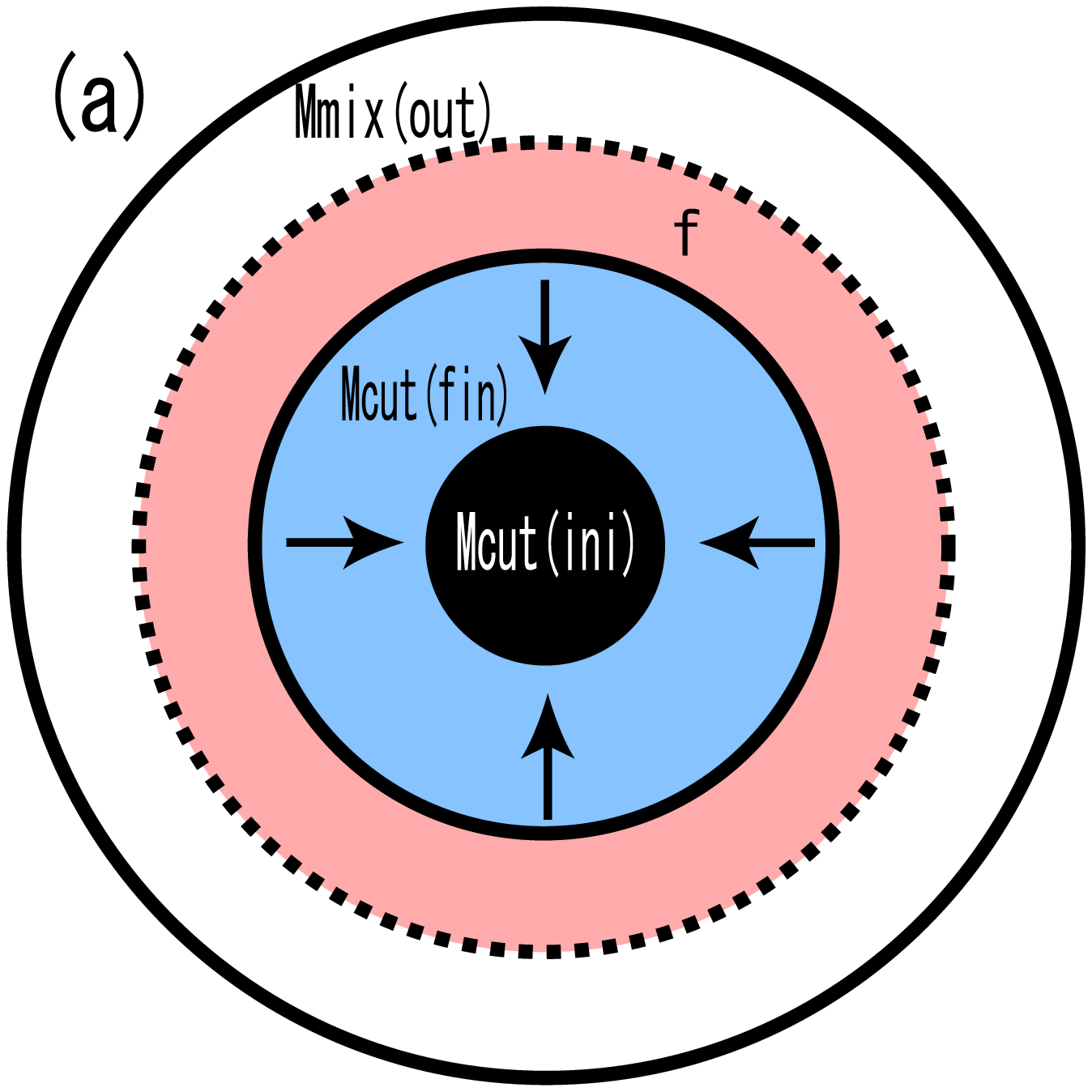}{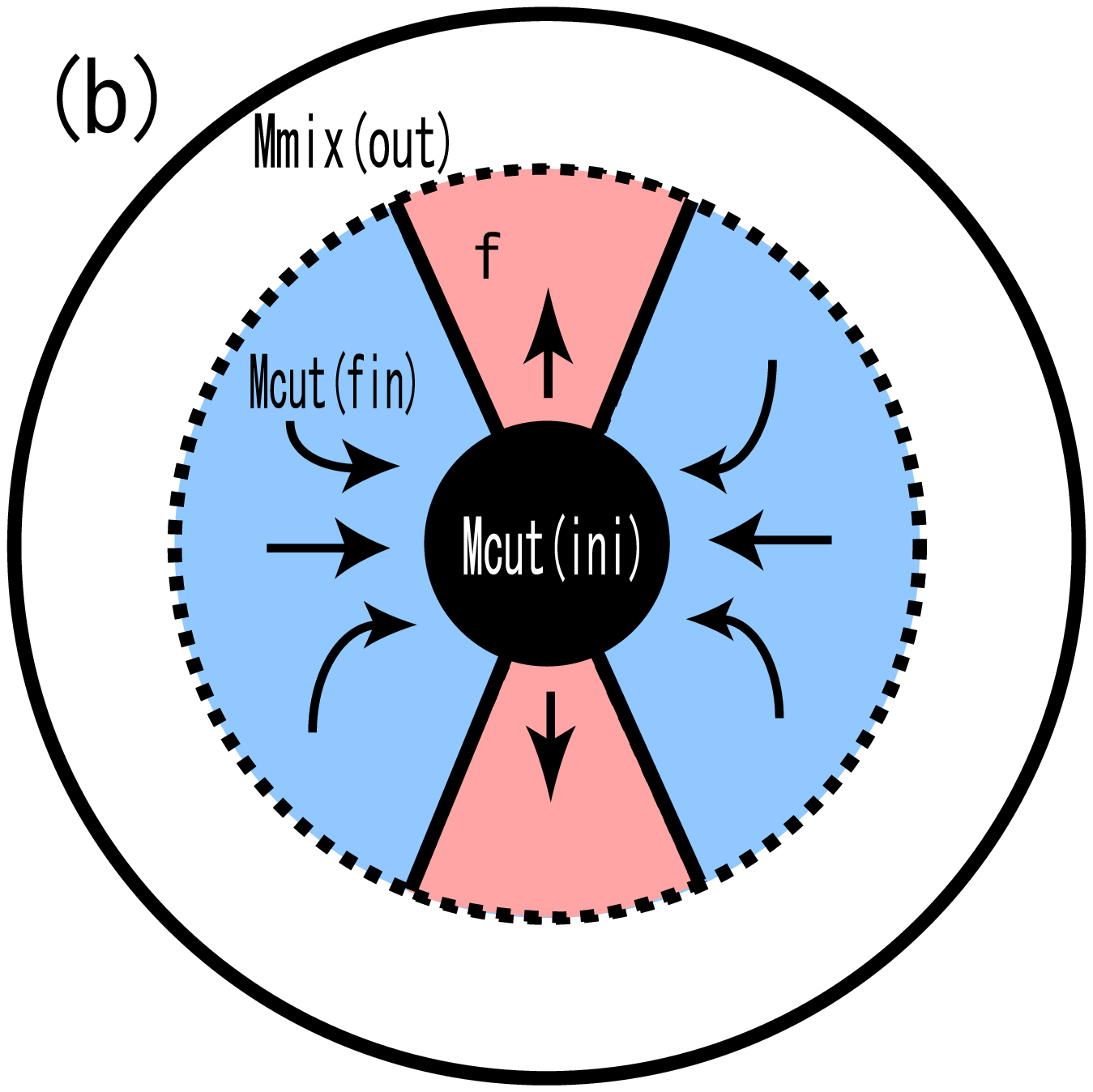}\figcaption{The illustration of the
mixing-fallback model. 
The central black region is the initial mass cut, that is, inside the inner
boundary of the mixing region, $M_{\rm cut}{\rm (ini)}$.
The mixing region is enclosed with the dotted line.
A fraction $f$ of the materials in the mixing region ejected to the
interstellar space. 
The rest materials, locating in the blue region, fallback into the central
remnant.
(a) 1-dimensional picture. The materials mixed up to a
given radius, and a part of the materials are ejected. 
(b) 2-dimensional picture. While the all materials in the outer
region above $M_{\rm mix}{\rm (out)}$, are ejected, the materials
in the mixing region may be ejected only along the jet-axis.
In the jet-like explosion, the ejection factor $f$ depends on the
jet-parameters (\eg an opening angle and an energy deposition rate).
\label{fig:MF}}
\end{figure*}

\begin{figure*}
\epsscale{2.1}
\plotone{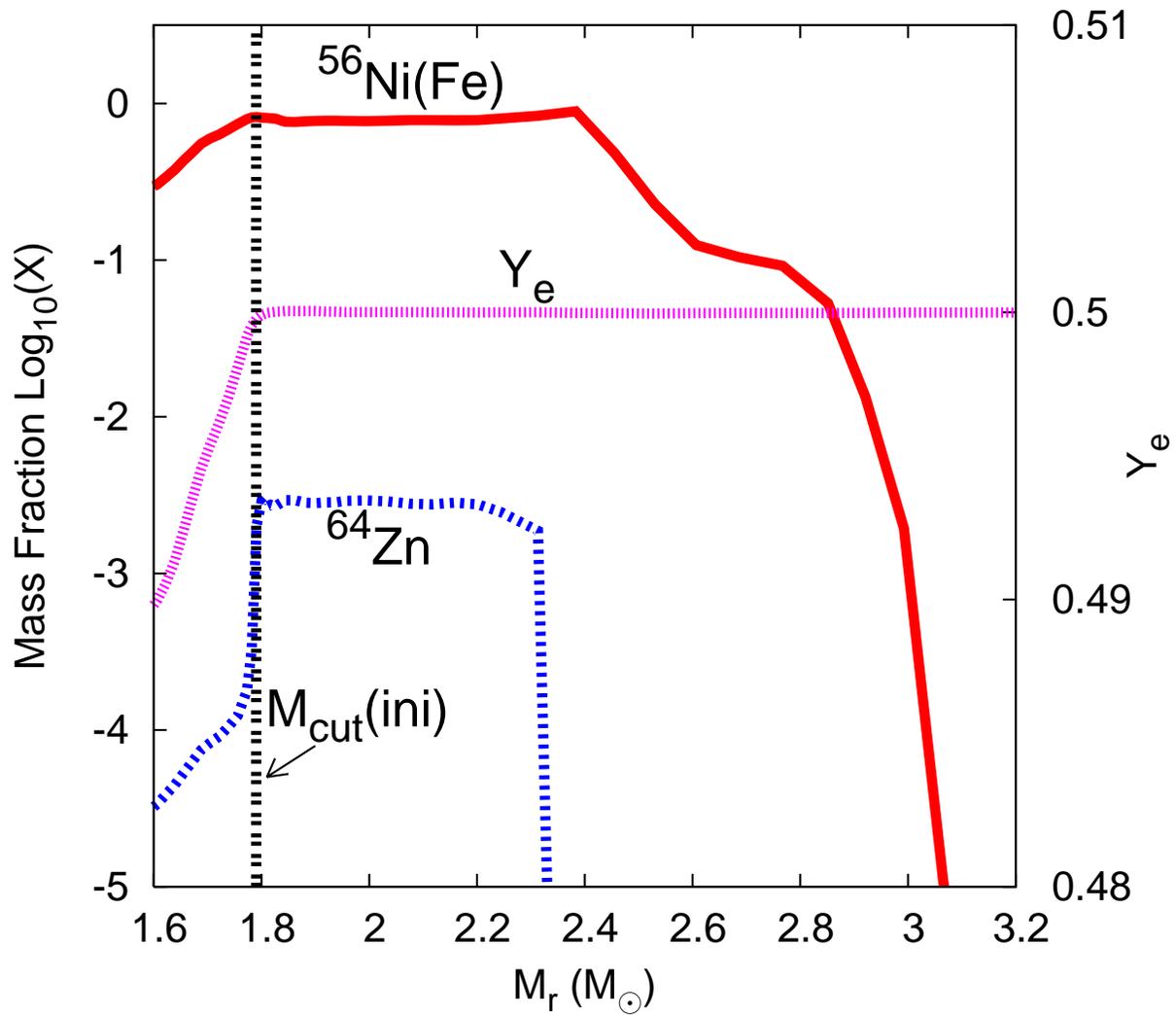} 
\figcaption{The abundance distribution (\Nifs;
{\it solid line}, ${\rm ^{64}Zn}$; {\it dashed line}) of the 25
$\Msun$, $E_{51}=10$ model around complete and incomplete Si burning
region. In the inner part ($M_r<1.8\Msun$)  
the Zn/Fe ratio is smaller 
because of lower $Y_{\rm e}$ ({\it dotted line}).  
$M_{\rm cut}{\rm (ini)}$ ({\it dotted and dashed line}) is
where $X({\rm ^{64}Zn})$ is smaller, \ie $Y_{\rm e} \lsim 0.5$, 
 as \cite{ume05}. \label{fig:25E20AD}}
\end{figure*}

\begin{figure*}
\plotone{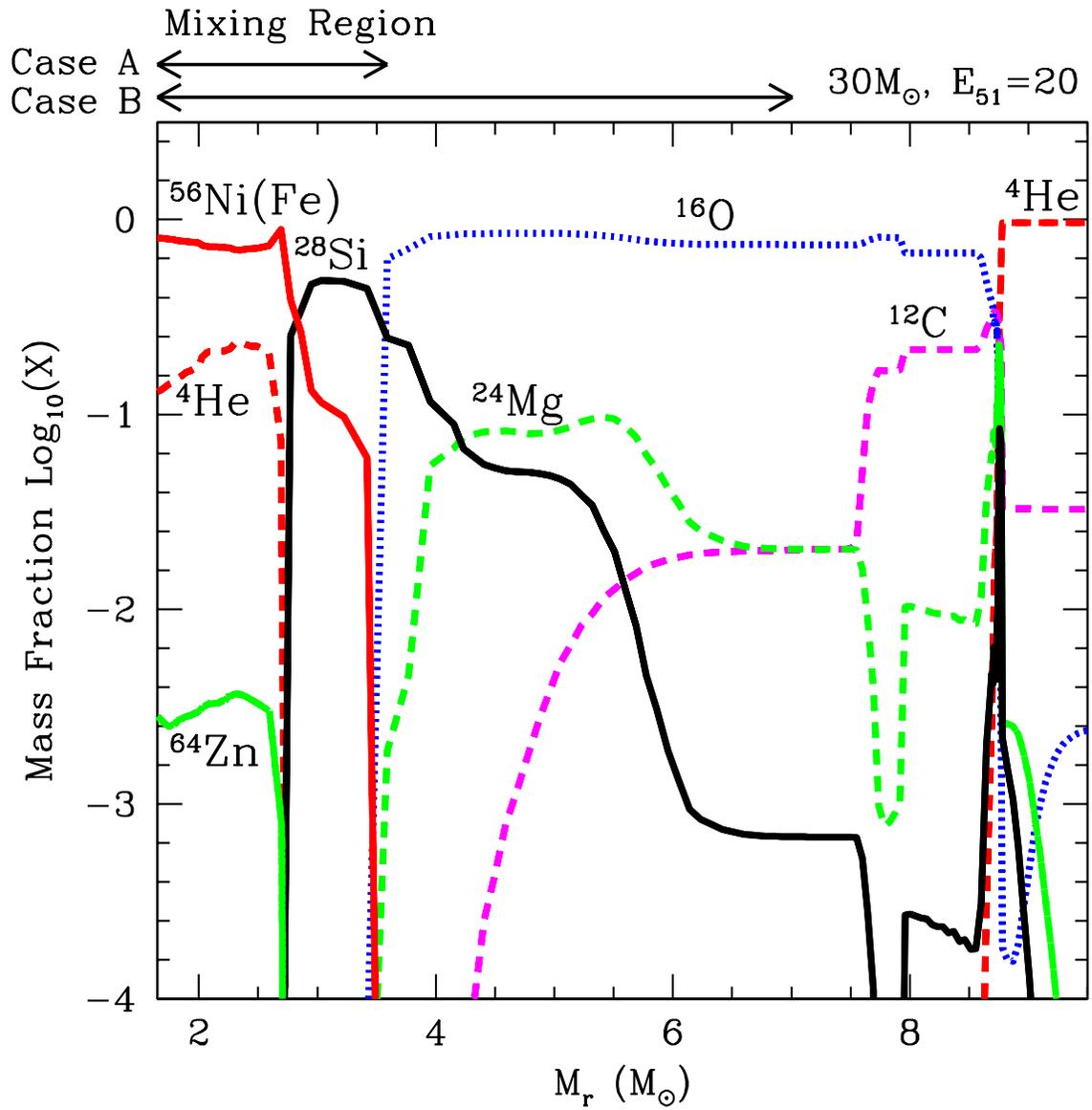} 
\figcaption{The abundance distribution of the 30
$\Msun$, $E_{51}=20$ model. 
The mixing regions of cases A and B are illustrated with arrows.
\label{fig:30E20AD}}
\end{figure*}

In the mixing-fallback model, the physical process is explained by three
parameters as follows:
\begin{itemize}
 \item $M_{\rm cut}{\rm (ini)}$: initial mass cut, which is
       corresponding to the inner boundary of the mixing region.
 \item $M_{\rm mix}{\rm (out)}$: outer boundary of the mixing region.
 \item $f$: a fraction of matter ejected from the mixed region. It
       determines [$\alpha$/Fe].
\end{itemize}
Figures~\ref{fig:MF}ab illustrate these parameters for spherical and
aspherical models. The final remnant mass, $M_{\rm cut}{\rm (fin)}$, is
determined by the above three parameters
\setcounter{equation}{6}
\begin{equation}
 M_{\rm cut}{\rm (fin)}=M_{\rm cut}{\rm (ini)}+(1-f)\times (M_{\rm mix}{\rm (out)} - M_{\rm cut}{\rm (ini)}).
\end{equation}
In this paper, we determine these parameters as follows.
\begin{itemize}
 \item $M_{\rm cut}{\rm (ini)}$: the initial mass cut is adopted so
       that [Zn/Fe] attains maximum, thus locates at the bottom of the
       $Y_{\rm e} \simeq 0.5$ layer (Fig.~\ref{fig:25E20AD}) 
       where is very close to the surface of Fe
       core of the progenitors. Figure~\ref{fig:25E20AD} shows the
       abundance distribution of the 25 $\Msun$ star with $E_{51}=10$
       around the complete Si burning region. For Population III SNe,
       the dominant isotope of Zn is $^{64}$Zn, which is the decay
       product of $^{64}$Ge. $^{64}$Zn is mostly produced in the
       complete Si burning region where $Y_{\rm e}\simeq 0.5$ and the Zn/Fe
       ratio decreases for lower $Y_{\rm e}$ (see Fig.~4 in
       \citealt{ume05}). Therefore, as the mass cut decreases, [Zn/Fe]
       in the ejecta first increases and then decreases. This choice of
       $M_{\rm cut}{\rm (ini)}$ tends to give a smaller estimate of
       $E_{51}$ to fit to the observed Zn/Fe, because Zn/Fe is larger
       for larger $E_{51}$.
 \item $M_{\rm mix}{\rm (out)}$ and $f$: we study two case as
       follows. Case B is an additional model for massive star models.
       The abundance distribution of the 30 $\Msun$, $E_{51}=20$ model
       is shown in Figure~\ref{fig:30E20AD} illustrating the mixing
       region in both cases.
\begin{description}
 \item[Case ``A'' :] We assume that the mixing occurs in the Si burning
	    region and thus $M_{\rm mix}{\rm (out)}$ is where
	    $X(^{56}{\rm Ni})=10^{-3}$. $f$ is chosen to yield [O/Fe]
	    $=0.5$. For most case, $f\sim 0.1$.
 \item[Case ``B'' :] For the $M\geq 30 \Msun$ models,
	    we consider another case. In this case, we assume that 
            $M_{\rm mix}{\rm (out)}$ is 2/3 of the O-rich layer, and
	    determine $f$ so that [Mg/Fe] $=0.2$ to be consistent with
	    the [Fe/H] $< -3.5$ stars in \cite{cay04}.
\end{description}
\end{itemize}

In this paper, we consider two cases A and B for $M\geq 30\Msun$
stars. 
[Mg/Fe] in case B is smaller than in case A but the ejected Fe mass in
case B is smaller than in case A (Tables~\ref{tab:models},
\ref{tab:modelsYe}, and \ref{tab:modelsYelow}). This is
because case B has larger amount of fallback, \ie larger
$M_{\rm mix}{\rm (out)}$ and smaller $f$, than case A. Such large
fallback which reaches the Mg-rich layer decreases the amount of ejected Mg 
and leads to a smaller Mg/Fe for case B than case A in spite of a smaller 
amount of ejected Fe. The observed HNe have variations of the
ejected Fe mass. Case A corresponds to typical HNe like SN~1998bw
that ejected $M(^{56}{\rm Ni})\sim 0.4\Msun$ \citep{nak01} and case B
corresponds to HNe like SN~1997ef that ejected
$M(^{56}{\rm Ni})\sim 0.15\Msun$ \citep{iwa00,maz00}.

Alternative interpretation of [Mg/Fe] $\sim 0.2$ is to eject a larger amount
of Fe than case A for the same amount of ejected Mg. In this
case, because of larger Fe, larger explosion energies 
($E_{51}\gsim 100$) are needed for this model to reproduce the small
[Fe/H] ($\sim -4$), as long as [Fe/H] is determined by Equation
(\ref{eq:SNSF}) (\ie the same constant $C$) of the SN-induced star
formation model. The explosion energies are considerably larger than
those of Pop I SNe. We thus do not consider such a possibility in this paper.

\clearpage

\headheight=-2.65cm
\textheight=21.4cm

\begin{deluxetable}{cccccccccccc}
\rotate
 \tabletypesize{\scriptsize}
 \tablecaption{The explosion models
 \label{tab:models}}
 \tablewidth{0pt}
 \tablehead{
   \colhead{$M_{\rm MS}$}
 & \colhead{$E_{51}$}
 & \colhead{$M({\rm ^{56}Ni})$}
 & \colhead{$M({\rm Mg})$}
 & \colhead{[Fe/H] \tablenotemark{a}}
 & \colhead{[O/Fe]}
 & \colhead{[Mg/Fe]}
 & \colhead{$M_{\rm cut}{\rm (fin)}$}
 & \colhead{$M_{\rm cut}{\rm (ini)}$}
 & \colhead{$M_{\rm mix}{\rm (out)}$}
 & \colhead{$f$ \tablenotemark{b}}
 & \colhead{The abundance pattern}
 }
\startdata
13  & 1   & 0.07   & 0.09 & $-$2.55 & $-$0.08 & 0.37 & 1.57  &      &       & 1     & Fig.~\ref{fig:VMP}a \\
15  & 1   & 0.07   & 0.07 & $-$2.55 &    0.15 & 0.27 & 1.48  &      &       & 1     & Fig.~\ref{fig:VMP}b \\
18  & 1   & 0.07   & 0.16 & $-$2.55 &    0.41 & 0.63 & 1.65  &      &       & 1     & Fig.~\ref{fig:VMP}c \\
20  & 10  & 0.08   & 0.17 & $-$3.34 &    0.50 & 0.58 & 1.90  & 1.52 & 2.05  & 0.28  & Fig.~\ref{fig:EMP}a \\
25  & 5   & 0.11   & 0.14 & $-$2.97 &    0.50 & 0.41 & 2.45  & 1.79 & 2.69  & 0.26  & Fig.~\ref{fig:25E5} \\
25  & 10  & 0.10   & 0.15 & $-$3.27 &    0.50 & 0.48 & 2.86  & 1.79 & 3.07  & 0.16  & Fig.~\ref{fig:EMP}b \\
25F & 1   & 0.0008 & 0.03 & $-$4.50 &    2.40 & 1.81 & 4.20  & 1.79 & 4.21  & 0.004 & Fig.~\ref{fig:faint} \\
30A & 20  & 0.16   & 0.22 & $-$3.31 &    0.50 & 0.41 & 3.27  & 1.65 & 3.59  & 0.17  & Fig.~\ref{fig:EMP}c \\
\enddata

\tablenotetext{a}{[Fe/H] is determined as [Fe/H] 
$\simeq \log_{10} (M({\rm Fe})/E_{51}^{6/7})-C$. Here we assume that $C$ is a
constant value, $C=1.4$. }
\tablenotetext{b}{$f$ is an ejection factor. $f=1$ corresponds to normal
 SN models without the mixing-fallback model. }

\tablecomments{The numbers shown are the main-sequence mass, the
 explosion energy, the ejected \Nifs\ mass, the ejected Mg mass, [Fe/H],
 [O/Fe], [Mg/Fe], the final central remnant mass, the mass of inner
 boundary of the mixing region, the mass of outer boundary of the mixing
 region, the ejection factor and the figure number of the comparison
 with the EMP and VMP stars. The masses are in unit of $\Msun$.
}
\end{deluxetable}

\headheight=8.35cm
\textheight=11.4cm

\begin{deluxetable}{crrrrrrrrrrrrr}
\rotate
 \tabletypesize{\scriptsize}
 \tablecaption{Yields of stable isotopes. \label{tab:yield}}
 \tablewidth{0pt}
 \tablehead{
   \colhead{Species}
 & \multicolumn{13}{c}{Yields [$\Msun$]} \\
   \colhead{Mass [$\Msun$]}
 & \colhead{13}
 & \colhead{15}
 & \colhead{18}
 & \colhead{20}
 & \colhead{25}
 & \colhead{25}
 & \colhead{25F}
 & \colhead{30A}
 & \colhead{30B}
 & \colhead{40A} 
 & \colhead{40B} 
 & \colhead{50A} 
 & \colhead{50B} \\
   \colhead{$E_{51}$}
 & \colhead{1}
 & \colhead{1}
 & \colhead{1}
 & \colhead{10}
 & \colhead{5}
 & \colhead{10}
 & \colhead{1}
 & \colhead{20}
 & \colhead{20}
 & \colhead{30} 
 & \colhead{30} 
 & \colhead{40} 
 & \colhead{40} 
 }
\startdata
p            &  6.59E+00  &  7.58E+00  &  8.43E+00  &  8.77E+00  &  1.06E+01  &  1.06E+01  &  1.06E+01  &  1.17E+01  &  1.17E+01  &  1.40E+01  &  1.40E+01  &  1.63E+01  &  1.63E+01\\
d            &  1.49E$-$16  &  1.69E$-$16  &  1.28E$-$16  &  8.66E$-$17  &  2.04E$-$16  &  2.06E$-$16  &  2.02E$-$16  &  1.09E$-$14  &  1.09E$-$14  &  1.66E$-$14  &  1.66E$-$14  &  1.24E$-$15  &  1.24E$-$15\\
$^{3}$He     &  4.12E$-$05  &  4.09E$-$05  &  3.33E$-$05  &  4.76E$-$05  &  2.11E$-$04  &  2.11E$-$04  &  2.11E$-$04  &  2.06E$-$04  &  2.06E$-$04  &  2.56E$-$05  &  2.56E$-$05  &  2.86E$-$05  &  2.86E$-$05\\
$^{4}$He     &  4.01E+00  &  4.40E+00  &  5.42E+00  &  5.96E+00  &  8.03E+00  &  8.02E+00  &  8.03E+00  &  9.54E+00  &  9.51E+00  &  1.18E+01  &  1.18E+01  &  1.56E+01  &  1.55E+01\\
$^{6}$Li     &  3.29E$-$23  &  1.28E$-$22  &  4.52E$-$23  &  1.36E$-$22  &  2.28E$-$20  &  4.26E$-$20  &  2.68E$-$21  &  3.50E$-$17  &  3.50E$-$17  &  5.39E$-$17  &  5.39E$-$17  &  2.02E$-$18  &  2.02E$-$18\\
$^{7}$Li     &  2.17E$-$10  &  2.94E$-$10  &  7.34E$-$11  &  2.79E$-$10  &  5.68E$-$09  &  5.68E$-$09  &  5.68E$-$09  &  2.36E$-$08  &  2.36E$-$08  &  3.42E$-$11  &  3.42E$-$11  &  8.78E$-$12  &  8.78E$-$12\\
$^{9}$Be     &  1.77E$-$20  &  3.22E$-$22  &  1.05E$-$22  &  4.83E$-$20  &  5.00E$-$17  &  3.69E$-$17  &  1.24E$-$17  &  3.09E$-$18  &  3.09E$-$18  &  9.03E$-$18  &  9.03E$-$18  &  6.04E$-$17  &  6.04E$-$17\\
$^{10}$B     &  3.16E$-$21  &  3.92E$-$20  &  2.86E$-$20  &  2.88E$-$19  &  2.79E$-$15  &  2.05E$-$13  &  2.67E$-$18  &  1.05E$-$14  &  1.05E$-$14  &  9.41E$-$15  &  9.41E$-$15  &  1.72E$-$17  &  7.85E$-$18\\
\enddata
\end{deluxetable}

\begin{deluxetable}{crrrrrrrrrrrrr}
\rotate
 \tabletypesize{\scriptsize}
 \tablecaption{Yields of unstable isotopes. \label{tab:yield2}}
 \tablewidth{0pt}
 \tablehead{
   \colhead{Species}
 & \multicolumn{13}{c}{Yields [$\Msun$]} \\
   \colhead{Mass [$\Msun$]}
 & \colhead{13}
 & \colhead{15}
 & \colhead{18}
 & \colhead{20}
 & \colhead{25}
 & \colhead{25}
 & \colhead{25F}
 & \colhead{30A}
 & \colhead{30B}
 & \colhead{40A} 
 & \colhead{40B} 
 & \colhead{50A} 
 & \colhead{50B} \\
   \colhead{$E_{51}$}
 & \colhead{1}
 & \colhead{1}
 & \colhead{1}
 & \colhead{10}
 & \colhead{5}
 & \colhead{10}
 & \colhead{1}
 & \colhead{20}
 & \colhead{20}
 & \colhead{30} 
 & \colhead{30} 
 & \colhead{40} 
 & \colhead{40} 
 }
\startdata
$^{22}$Na    &  1.51E$-$07  &  7.70E$-$06  &  1.68E$-$05  &  4.65E$-$05  &  9.97E$-$06  &  6.54E$-$06  &  8.98E$-$06  &  2.70E$-$05  &  1.34E$-$05  &  1.37E$-$05  &  1.19E$-$05  &  1.77E$-$04  &  9.12E$-$05\\
$^{26}$Al    &  9.80E$-$07  &  1.10E$-$06  &  3.32E$-$06  &  8.93E$-$07  &  1.21E$-$06  &  1.26E$-$06  &  5.34E$-$09  &  2.87E$-$06  &  1.32E$-$06  &  2.13E$-$05  &  1.76E$-$05  &  5.34E$-$06  &  1.32E$-$06\\
$^{41}$Ca    &  3.72E$-$07  &  7.42E$-$07  &  1.51E$-$06  &  2.63E$-$06  &  5.94E$-$06  &  4.00E$-$06  &  7.98E$-$08  &  3.63E$-$06  &  2.04E$-$07  &  2.12E$-$05  &  1.14E$-$06  &  1.72E$-$06  &  4.78E$-$07\\
$^{44}$Ti    &  1.70E$-$05  &  2.17E$-$05  &  1.40E$-$05  &  1.25E$-$04  &  4.26E$-$05  &  6.82E$-$05  &  1.82E$-$07  &  1.80E$-$04  &  5.67E$-$05  &  1.76E$-$04  &  7.16E$-$05  &  6.08E$-$04  &  3.95E$-$04\\
$^{60}$Fe    &  4.47E$-$16  &  2.44E$-$14  &  4.24E$-$15  &  1.10E$-$13  &  1.54E$-$13  &  1.10E$-$12  &  1.01E$-$15  &  2.10E$-$12  &  1.95E$-$12  &  2.44E$-$11  &  2.44E$-$11  &  1.11E$-$12  &  3.86E$-$13\\
$^{56}$Ni    &  7.00E$-$02  &  7.00E$-$02  &  7.00E$-$02  &  8.30E$-$02  &  1.07E$-$01  &  9.70E$-$02  &  7.93E$-$04  &  1.60E$-$01  &  5.04E$-$02  &  2.58E$-$01  &  1.08E$-$01  &  3.61E$-$01  &  2.34E$-$01\\
\enddata
\end{deluxetable}

\headheight=-1.65cm
\textheight=21.4cm

\begin{deluxetable}{crrrrrr}
 \tabletypesize{\scriptsize}
 \tablecaption{IMF-integrated Yields. \label{tab:IMF}}
 \tablewidth{0pt}
 \tablehead{
   \colhead{Species}
 & \multicolumn{6}{c}{Yields [Mass fractions]} \\
   \colhead{$\Mms\geq30\Msun$ models}
 & \colhead{A}
 & \colhead{B}
 & \colhead{A} 
 & \colhead{B} 
 & \colhead{A} 
 & \colhead{B} \\
   \colhead{``n/p ratio''}
 & \colhead{No}
 & \colhead{No}
 & \colhead{Yes} 
 & \colhead{Yes} 
 & \colhead{Yes} 
 & \colhead{Yes} \\
   \colhead{``low-density''}
 & \colhead{No}
 & \colhead{No}
 & \colhead{No} 
 & \colhead{No} 
 & \colhead{Yes} 
 & \colhead{Yes} \\
   \colhead{The abundance pattern} 
 & \multicolumn{2}{c}{Fig.~\ref{fig:IMF}}
 & \multicolumn{2}{c}{Fig.~\ref{fig:Ye}j} 
 & \multicolumn{2}{c}{Fig.~\ref{fig:Yelow}g} 
 }
\startdata
p            &  3.28E$-$02  &  3.28E$-$02  &  3.27E$-$02  &  3.27E$-$02  &  3.28E$-$02  &  3.28E$-$02\\
d            &  8.97E$-$18  &  8.97E$-$18  &  8.97E$-$18  &  8.97E$-$18  &  1.22E$-$15  &  1.22E$-$15\\
$^{3}$He     &  2.72E$-$07  &  2.72E$-$07  &  2.72E$-$07  &  2.72E$-$07  &  2.72E$-$07  &  2.72E$-$07\\
$^{4}$He     &  2.31E$-$02  &  2.31E$-$02  &  2.30E$-$02  &  2.30E$-$02  &  2.33E$-$02  &  2.32E$-$02\\
$^{6}$Li     &  2.75E$-$20  &  2.75E$-$20  &  2.75E$-$20  &  2.75E$-$20  &  4.04E$-$18  &  4.04E$-$18\\
$^{7}$Li     &  1.14E$-$11  &  1.14E$-$11  &  1.12E$-$11  &  1.12E$-$11  &  1.14E$-$11  &  1.14E$-$11\\
$^{9}$Be     &  2.40E$-$20  &  2.40E$-$20  &  2.50E$-$20  &  2.50E$-$20  &  2.09E$-$19  &  2.09E$-$19\\
$^{10}$B     &  8.94E$-$17  &  8.94E$-$17  &  4.80E$-$17  &  4.80E$-$17  &  7.70E$-$18  &  7.70E$-$18\\
\enddata
\end{deluxetable}

\begin{deluxetable}{cccccccccccc}
\rotate
 \tabletypesize{\scriptsize}
 \tablecaption{The explosion models with the $Y_{\rm e}$ modification.
 \label{tab:modelsYe}}
 \tablewidth{0pt}
 \tablehead{
   \colhead{$M_{\rm MS}$}
 & \colhead{$E_{51}$}
 & \colhead{$M({\rm ^{56}Ni})$}
 & \colhead{$M({\rm Mg})$}
 & \colhead{[Fe/H] \tablenotemark{a}}
 & \colhead{[O/Fe]}
 & \colhead{[Mg/Fe]}
 & \colhead{$M_{\rm cut}{\rm (fin)}$}
 & \colhead{$M_{\rm cut}{\rm (ini)}$}
 & \colhead{$M_{\rm mix}{\rm (out)}$}
 & \colhead{$f$ \tablenotemark{b}}
 & \colhead{The abundance pattern}
 }
\startdata
13  & 1   & 0.07   & 0.09 & $-$2.55 & $-$0.08 & 0.37 & 1.57  &      &       & 1    & Fig.~\ref{fig:Ye}a \\
15  & 1   & 0.07   & 0.07 & $-$2.55 &    0.15 & 0.27 & 1.48  &      &       & 1    & Fig.~\ref{fig:Ye}b \\
18  & 1   & 0.07   & 0.16 & $-$2.55 &    0.40 & 0.63 & 1.65  &      &       & 1    & Fig.~\ref{fig:Ye}c \\
20  & 10  & 0.08   & 0.17 & $-$3.34 &    0.50 & 0.58 & 1.88  & 1.52 & 2.02  & 0.28 & Fig.~\ref{fig:Ye}d \\
25  & 5   & 0.10   & 0.14 & $-$2.98 &    0.50 & 0.41 & 2.39  & 1.79 & 2.61  & 0.27 & Fig.~\ref{fig:Ye}i \\
25  & 10  & 0.10   & 0.16 & $-$3.25 &    0.50 & 0.48 & 2.90  & 1.79 & 2.99  & 0.07 & Fig.~\ref{fig:Ye}e \\
30A & 20  & 0.16   & 0.22 & $-$3.31 &    0.50 & 0.41 & 3.27  & 1.65 & 3.59  & 0.17 & Fig.~\ref{fig:Ye}f \\
30B & 20  & 0.05   & 0.04 & $-$3.82 &    0.55 & 0.20 & 6.73  & 1.65 & 7.01  & 0.05 & Fig.~\ref{fig:Ye}f \\
\enddata

\tablenotetext{a}{[Fe/H] is determined as [Fe/H] 
$\simeq \log_{10} (M({\rm Fe})/E_{51}^{6/7})-C$. Here we assume that $C$ is a
constant value, $C=1.4$. }
\tablenotetext{b}{$f$ is an ejection factor. $f=1$ corresponds to normal
 SN models without the mixing-fallback model. }
\tablecomments{Same as Table~\ref{tab:models}, but for models applied
 the $Y_{\rm e}$ modification described in \S~\ref{sec:improve}.}
\end{deluxetable}

\headheight=6.35cm
\textheight=13.4cm

\begin{deluxetable}{crrrrrrrrrrrr}
\rotate
 \tabletypesize{\scriptsize}
 \tablecaption{Yields of stable isotopes for the models with the $Y_{\rm e}$ modification. \label{tab:yieldYe}}
 \tablewidth{0pt}
 \tablehead{
   \colhead{Species}
 & \multicolumn{12}{c}{Yields [$\Msun$]} \\
   \colhead{Mass [$\Msun$]}
 & \colhead{13}
 & \colhead{15}
 & \colhead{18}
 & \colhead{20}
 & \colhead{25}
 & \colhead{25}
 & \colhead{30A}
 & \colhead{30B}
 & \colhead{40A} 
 & \colhead{40B} 
 & \colhead{50A} 
 & \colhead{50B} \\
   \colhead{$E_{51}$}
 & \colhead{1}
 & \colhead{1}
 & \colhead{1}
 & \colhead{10}
 & \colhead{5}
 & \colhead{10}
 & \colhead{20}
 & \colhead{20}
 & \colhead{30} 
 & \colhead{30} 
 & \colhead{40} 
 & \colhead{40} 
 }
\startdata
p            &  6.59E+00  &  7.58E+00  &  8.43E+00  &  8.77E+00  &  1.06E+01  &  1.05E+01  &  1.17E+01  &  1.17E+01  &  1.40E+01  &  1.40E+01  &  1.63E+01  &  1.63E+01\\
d            &  1.49E$-$16  &  1.69E$-$16  &  1.28E$-$16  &  8.66E$-$17  &  2.04E$-$16  &  2.06E$-$16  &  1.09E$-$14  &  1.09E$-$14  &  1.66E$-$14  &  1.66E$-$14  &  1.24E$-$15  &  1.24E$-$15\\
$^{3}$He     &  4.12E$-$05  &  4.09E$-$05  &  3.33E$-$05  &  4.76E$-$05  &  2.11E$-$04  &  2.11E$-$04  &  2.06E$-$04  &  2.06E$-$04  &  2.56E$-$05  &  2.56E$-$05  &  2.86E$-$05  &  2.86E$-$05\\
$^{4}$He     &  4.01E+00  &  4.40E+00  &  5.42E+00  &  5.96E+00  &  8.03E+00  &  7.82E+00  &  9.54E+00  &  9.51E+00  &  1.18E+01  &  1.18E+01  &  1.56E+01  &  1.55E+01\\
$^{6}$Li     &  3.65E$-$23  &  1.11E$-$22  &  4.37E$-$23  &  1.36E$-$22  &  2.28E$-$20  &  2.72E$-$20  &  3.50E$-$17  &  3.50E$-$17  &  5.39E$-$17  &  5.39E$-$17  &  2.02E$-$18  &  2.02E$-$18\\
$^{7}$Li     &  2.17E$-$10  &  2.94E$-$10  &  7.34E$-$11  &  2.79E$-$10  &  5.68E$-$09  &  5.12E$-$09  &  2.36E$-$08  &  2.36E$-$08  &  3.42E$-$11  &  3.42E$-$11  &  8.78E$-$12  &  8.78E$-$12\\
$^{9}$Be     &  1.77E$-$20  &  3.22E$-$22  &  1.05E$-$22  &  4.83E$-$20  &  5.00E$-$17  &  3.96E$-$17  &  3.09E$-$18  &  3.09E$-$18  &  9.03E$-$18  &  9.03E$-$18  &  6.04E$-$17  &  6.04E$-$17\\
$^{10}$B     &  2.92E$-$21  &  8.30E$-$20  &  3.92E$-$21  &  2.88E$-$19  &  2.79E$-$15  &  8.28E$-$14  &  2.95E$-$14  &  2.95E$-$14  &  1.41E$-$14  &  1.41E$-$14  &  1.51E$-$17  &  5.27E$-$18\\
\enddata
\end{deluxetable}

\begin{deluxetable}{crrrrrrrrrrrr}
\rotate
 \tabletypesize{\scriptsize}
 \tablecaption{Yields of unstable isotopes for the models with the $Y_{\rm e}$ modification. \label{tab:yield2Ye}}
 \tablewidth{0pt}
 \tablehead{
   \colhead{Species}
 & \multicolumn{12}{c}{Yields [$\Msun$]} \\
   \colhead{Mass [$\Msun$]}
 & \colhead{13}
 & \colhead{15}
 & \colhead{18}
 & \colhead{20}
 & \colhead{25}
 & \colhead{25}
 & \colhead{30A}
 & \colhead{30B}
 & \colhead{40A} 
 & \colhead{40B} 
 & \colhead{50A} 
 & \colhead{50B} \\
   \colhead{$E_{51}$}
 & \colhead{1}
 & \colhead{1}
 & \colhead{1}
 & \colhead{10}
 & \colhead{5}
 & \colhead{10}
 & \colhead{20}
 & \colhead{20}
 & \colhead{30} 
 & \colhead{30} 
 & \colhead{40} 
 & \colhead{40} 
 }
\startdata
$^{22}$Na    &  1.51E$-$07  &  7.70E$-$06  &  1.68E$-$05  &  4.65E$-$05  &  9.97E$-$06  &  7.01E$-$06  &  2.70E$-$05  &  1.34E$-$05  &  1.37E$-$05  &  1.19E$-$05  &  1.77E$-$04  &  9.14E$-$05\\
$^{26}$Al    &  9.84E$-$07  &  1.12E$-$06  &  3.33E$-$06  &  9.81E$-$07  &  1.21E$-$06  &  1.37E$-$06  &  2.92E$-$06  &  1.33E$-$06  &  2.13E$-$05  &  1.76E$-$05  &  5.24E$-$06  &  1.26E$-$06\\
$^{41}$Ca    &  3.54E$-$07  &  8.05E$-$07  &  1.72E$-$06  &  3.73E$-$06  &  6.75E$-$06  &  6.94E$-$06  &  3.66E$-$06  &  2.15E$-$07  &  2.13E$-$05  &  1.20E$-$06  &  1.43E$-$05  &  2.82E$-$06\\
$^{44}$Ti    &  9.55E$-$06  &  1.19E$-$05  &  8.75E$-$06  &  1.09E$-$04  &  2.86E$-$05  &  5.25E$-$05  &  1.49E$-$04  &  4.63E$-$05  &  1.53E$-$04  &  6.28E$-$05  &  5.37E$-$04  &  3.51E$-$04\\
$^{60}$Fe    &  4.74E$-$16  &  7.17E$-$14  &  3.78E$-$15  &  1.09E$-$13  &  1.53E$-$13  &  2.37E$-$12  &  1.68E$-$12  &  1.11E$-$12  &  2.40E$-$11  &  2.40E$-$11  &  7.78E$-$13  &  1.45E$-$13\\
$^{56}$Ni    &  7.00E$-$02  &  7.00E$-$02  &  7.00E$-$02  &  8.27E$-$02  &  1.04E$-$01  &  1.03E$-$01  &  1.60E$-$01  &  4.97E$-$02  &  2.56E$-$01  &  1.08E$-$01  &  3.59E$-$01  &  2.35E$-$01\\
\enddata
\end{deluxetable}

\headheight=-1.65cm
\textheight=21.4cm

\begin{deluxetable}{cccccccccccccccc}
\rotate
 \tabletypesize{\scriptsize}
 \tablecaption{The explosion models with the $Y_{\rm e}$ and
 ``low-density'' modifications.
 \label{tab:modelsYelow}}
 \tablewidth{0pt}
 \tablehead{
   \colhead{$M_{\rm MS}$}
 & \colhead{$E_{51}$}
 & \colhead{$M({\rm ^{56}Ni})$}
 & \colhead{$M({\rm Mg})$}
 & \colhead{[Fe/H] \tablenotemark{a}}
 & \colhead{[O/Fe]}
 & \colhead{[Mg/Fe]}
 & \colhead{$M_{\rm cut}{\rm (fin)}$}
 & \colhead{$M_{\rm cut}{\rm (ini)}$}
 & \colhead{$M_{\rm mix}{\rm (out)}$}
 & \colhead{$f$ \tablenotemark{b}}
 & \colhead{``low-density'' factor}
 & \colhead{The abundance pattern}
 }
\startdata
20  & 10  & 0.08   & 0.16 & $-$3.33 &    0.50 & 0.56 & 1.83  & 1.52 & 2.01  & 0.37 & 1/2 & Fig.~\ref{fig:Yelow}a \\
25  & 5   & 0.11   & 0.14 & $-$2.95 &    0.50 & 0.35 & 2.08  & 1.79 & 2.38  & 0.51 & 1/4 & Fig.~\ref{fig:Yelow}f \\
25  & 10  & 0.10   & 0.14 & $-$3.25 &    0.50 & 0.41 & 2.67  & 1.79 & 2.92  & 0.22 & 1/3 & Fig.~\ref{fig:Yelow}b \\
30A & 20  & 0.16   & 0.21 & $-$3.31 &    0.50 & 0.38 & 2.97  & 1.65 & 3.42  & 0.25 & 1/3 & Fig.~\ref{fig:Yelow}c \\
30B & 20  & 0.05   & 0.04 & $-$3.86 &    0.64 & 0.20 & 6.50  & 1.65 & 6.87  & 0.07 & 1/3 & Fig.~\ref{fig:Yelow}c \\
40A & 30  & 0.28   & 0.38 & $-$3.22 &    0.50 & 0.40 & 4.80  & 2.24 & 5.53  & 0.22 & 1/4 & Fig.~\ref{fig:Yelow}d \\
40B & 30  & 0.12   & 0.10 & $-$3.59 &    0.49 & 0.20 & 10.09 & 2.24 & 10.92 & 0.10 & 1/4 & Fig.~\ref{fig:Yelow}d \\
50A & 40  & 0.37   & 0.51 & $-$3.20 &    0.50 & 0.42 & 3.18  & 1.89 & 3.99  & 0.39 & 1/2 & Fig.~\ref{fig:Yelow}e  \\
\enddata
\tablenotetext{a}{[Fe/H] is determined as [Fe/H] 
$\simeq \log_{10} (M({\rm Fe})/E_{51}^{6/7})-C$. Here we assume that $C$ is a
constant value, $C=1.4$. }
\tablenotetext{b}{$f$ is an ejection factor.}
\tablecomments{Same as Table~\ref{tab:models}, but for models applied
 the $Y_{\rm e}$ and ``low-density'' modifications described in \S~\ref{sec:improve}.}
\end{deluxetable}

\headheight=-1.65cm
\textheight=21.4cm

\begin{deluxetable}{crrrrrrrrr}
 \tabletypesize{\scriptsize}
 \tablecaption{Yields of stable isotopes for the models with the $Y_{\rm e}$ and
 ``low-density'' modifications. \label{tab:yieldYelow}}
 \tablewidth{0pt}
 \tablehead{
   \colhead{Species}
 & \multicolumn{9}{c}{Yields [$\Msun$]} \\
   \colhead{Mass [$\Msun$]}
 & \colhead{20}
 & \colhead{25}
 & \colhead{25}
 & \colhead{30A}
 & \colhead{30B}
 & \colhead{40A} 
 & \colhead{40B} 
 & \colhead{50A} 
 & \colhead{50B} \\
   \colhead{$E_{51}$}
 & \colhead{10}
 & \colhead{5}
 & \colhead{10}
 & \colhead{20}
 & \colhead{20}
 & \colhead{30} 
 & \colhead{30} 
 & \colhead{40} 
 & \colhead{40} 
 }
\startdata
p            &  8.77E+00  &  1.06E+01  &  1.06E+01  &  1.17E+01  &  1.17E+01  &  1.40E+01  &  1.40E+01  &  1.63E+01  &  1.63E+01\\
d            &  8.66E$-$17  &  2.10E$-$16  &  2.17E$-$16  &  4.19E$-$14  &  4.19E$-$14  &  4.43E$-$12  &  4.43E$-$12  &  1.23E$-$15  &  1.23E$-$15\\
$^{3}$He     &  4.76E$-$05  &  2.11E$-$04  &  2.11E$-$04  &  2.06E$-$04  &  2.06E$-$04  &  2.56E$-$05  &  2.56E$-$05  &  2.86E$-$05  &  2.86E$-$05\\
$^{4}$He     &  5.98E+00  &  8.07E+00  &  8.06E+00  &  9.64E+00  &  9.55E+00  &  1.21E+01  &  1.20E+01  &  1.57E+01  &  1.57E+01\\
$^{6}$Li     &  1.22E$-$22  &  3.06E$-$20  &  7.78E$-$20  &  1.37E$-$16  &  1.37E$-$16  &  1.47E$-$14  &  1.47E$-$14  &  2.00E$-$18  &  2.00E$-$18\\
$^{7}$Li     &  2.79E$-$10  &  5.68E$-$09  &  5.68E$-$09  &  2.36E$-$08  &  2.36E$-$08  &  3.50E$-$11  &  3.50E$-$11  &  8.78E$-$12  &  8.78E$-$12\\
$^{9}$Be     &  1.16E$-$20  &  3.12E$-$17  &  2.90E$-$17  &  6.90E$-$18  &  6.90E$-$18  &  6.87E$-$16  &  6.87E$-$16  &  9.43E$-$17  &  9.43E$-$17\\
$^{10}$B     &  2.59E$-$19  &  1.95E$-$15  &  3.81E$-$15  &  5.17E$-$16  &  5.17E$-$16  &  2.20E$-$14  &  2.20E$-$14  &  9.04E$-$18  &  4.50E$-$18\\
\enddata
\end{deluxetable}

\begin{deluxetable}{crrrrrrrrr}
 \tabletypesize{\scriptsize}
 \tablecaption{Yields of unstable isotopes for the models with the $Y_{\rm e}$ and
 ``low-density'' modifications. \label{tab:yield2Yelow}}
 \tablewidth{0pt}
 \tablehead{
   \colhead{Species}
 & \multicolumn{9}{c}{Yields [$\Msun$]} \\
   \colhead{Mass [$\Msun$]}
 & \colhead{20}
 & \colhead{25}
 & \colhead{25} 
 & \colhead{30A}
 & \colhead{30B}
 & \colhead{40A} 
 & \colhead{40B} 
 & \colhead{50A} 
 & \colhead{50B} \\
   \colhead{$E_{51}$}
 & \colhead{10}
 & \colhead{5}
 & \colhead{10}
 & \colhead{20}
 & \colhead{20}
 & \colhead{30} 
 & \colhead{30} 
 & \colhead{40} 
 & \colhead{40} 
 }
\startdata
$^{22}$Na    &  4.74E$-$05  &  1.09E$-$05  &  6.81E$-$06  &  3.00E$-$05  &  1.54E$-$05  &  4.11E$-$06  &  3.33E$-$06  &  1.82E$-$04  &  1.06E$-$04\\
$^{26}$Al    &  8.17E$-$07  &  9.13E$-$07  &  1.04E$-$06  &  2.57E$-$06  &  9.78E$-$07  &  4.26E$-$06  &  1.22E$-$06  &  7.45E$-$06  &  3.74E$-$06\\
$^{41}$Ca    &  3.40E$-$06  &  1.06E$-$05  &  3.69E$-$06  &  7.94E$-$06  &  6.26E$-$07  &  1.30E$-$05  &  1.37E$-$06  &  1.62E$-$05  &  4.52E$-$06\\
$^{44}$Ti    &  2.05E$-$04  &  1.68E$-$04  &  1.51E$-$04  &  4.93E$-$04  &  1.37E$-$04  &  8.70E$-$04  &  3.69E$-$04  &  1.26E$-$03  &  8.82E$-$04\\
$^{60}$Fe    &  2.45E$-$13  &  2.25E$-$13  &  9.60E$-$13  &  5.11E$-$13  &  2.34E$-$13  &  2.74E$-$12  &  2.74E$-$12  &  1.69E$-$12  &  4.60E$-$13\\
$^{56}$Ni    &  8.41E$-$02  &  1.12E$-$01  &  1.01E$-$01  &  1.62E$-$01  &  4.51E$-$02  &  2.81E$-$01  &  1.20E$-$01  &  3.71E$-$01  &  2.61E$-$01\\
$^{57}$Ni    &  1.27E$-$03  &  1.46E$-$03  &  1.38E$-$03  &  2.66E$-$03  &  7.43E$-$04  &  4.27E$-$03  &  1.81E$-$03  &  6.22E$-$03  &  4.37E$-$03
\enddata
\end{deluxetable}

\clearpage

\headheight=-1.65cm
\textheight=21.4cm

\clearpage
\clearpage

\begin{thebibliography}{}

\bibitem[Aoki \etal (2002)]{aok02} Aoki, W., Norris, J. E., Ryan,
			   S. G., Beers, T. C., \& Ando, H.
			   2002, PASJ, 54, 933

\bibitem[Aoki \etal (2004)]{aok04} Aoki, W., Norris, J. E., Ryan,
			   S. G., Beers, T. C., Christlieb,
			   N., Tsangarides, S., \& Ando, H.
			   2004, \apj, 608, 971

\bibitem[Anders \& Grevesse (1989)]{and89} Anders, E., \& Grevesse,
			   N. 1989, Geochim. Cosmochim. Acta, 53, 197

\bibitem[Argast \etal(2000)]{arg00} Argast, D., Samland, M., Gerhard,
			   O. E., \&  Thielemann, F.-K. 2000, \aap, 356,
			   873

\bibitem[Argast \etal(2002)]{arg02} Argast, D., Samland, M., Thielemann,
			   F.-K., \& Gerhard, O. E. 2002, \aap, 388, 842

\bibitem[Argast \etal(2004)]{arg04} Argast, D., Samland, M., Thielemann,
			   F.-K., \& Qian, Y.-Z. 2004, \aap, 416, 997

\bibitem[Aufderheide \etal(1991)]{auf91} Aufderheide,
			   M. B., Baron, E., \& Thielemann, F.-K. 1991,
			   \apj, 370, 630

\bibitem[Audouze \& Silk (1995)]{aud95} Audouze, J., \& Silk, J. 1995,
				\apj, 451, L49

\bibitem[Beers \& Christlieb(2005)]{bee05} Beers, T.C., \& Christlieb,
			   N. 2005, \araa, 43, 531

\bibitem[Bensby et al.(2003)]{ben03}
Bensby T., Feltzing S., Lundstr\"{o}m I. 2003, \aap, 410, 527

\bibitem[Blinnikov \etal(2000)]{bli00} Blinnikov, S., Lundqvist, P.,
			   Bartunov, O., Nomoto, K., \& Iwamoto,
			   K. 2000, \apj, 532, 1132

\bibitem[Bromm \& Larson (2004)]{bro04} Bromm, V., \& Larson, R. B. 2004,
				\araa, 42, 79

\bibitem[Buras \etal (2006a)]{bur06a} Buras, R., Janka, H.-Th., Rampp,
			     M., \& Kifonidis, K. 2006a, \aap, 457, 281
			   
\bibitem[Buras \etal (2006b)]{bur06b} Buras, R., Rampp, M., Janka,
			   H.-Th., \& Kifonidis, K. 2006b, \aap, 447,
			   1049

\bibitem[Burris \etal(2000)]{bur00} Burris, D. L., Pilachowski, C. A.,
			   Armandroff, T. E., Sneden, C., Cowan, J. J.,
			   \& Roe, H. 2000, \apj, 544, 302

\bibitem[Carretta et al.(2000)]{car00}
Carretta, E., Gratton, R. G., Sneden, C. 2000, \aap, 356, 238

\bibitem[Caughlan \etal(1985)]{cau85} Caughlan, G. R., Fowler,
				W. A., Harris, M. J., \& Zimmermann,
			   B. A. 1985, At Data Nucl. Data Tables, 32, 197

\bibitem[Caughlan \& Fowler (1988)]{cau88} Caughlan, G. R., \& Fowler,
				W. A. 1988, At Data Nucl. Data Tables,
				40, 283

\bibitem[CA04()]{cay04}Cayrel, R., \etal\  2004,
			   \aap, 416, 1117 (CA04)

\bibitem[CL02()]{chi02} Chieffi, A., \& Limongi, M. 2002,
\apj, 577, 281 (CL02)

\bibitem[Christlieb \etal (2002)]{chr02}Christlieb, N., \etal\ 2002, \nat,
			   419, 904

\bibitem[Chugai \etal (2005)]{chu05}Chugai, N.N., Fabrika, S.N.,
			   Sholukhova, O.N., Goranskij, V.P., Abolmasov,
			   P.K., \& Vlasyuk, V.V. 2005, Astron. Lett.,
			   31, 792

\bibitem[Cioffi \etal (1988)]{cio88} Cioffi, D. F.,
			     McKee, C. F.,
			     Bertschinger, E. 1988, \apj, 334, 252

\bibitem[Cobb \etal(2004)]{cob04} Cobb, B. E., Bailyn, C. D., van
			   Dokkum, P. G., Buxton, M. M., \& Bloom,
			   J. S. 2004, \apj, 608, L93

\bibitem[Coc \etal(2004)]{coc04} Coc, A., Vangioni-Flam, E.,
			   Descouvemont, P., Adahchour, A., \& Angulo,
			   C. 2004, \apj, 600, 544

\bibitem[Cohen \etal(2006)]{coh06} Cohen, J.G., \etal\ 2006, \aj, 132, 137

\bibitem[Deng \etal(2005)]{den05} Deng, J., Tominaga, N., Mazzali, P.A.,
			   Maeda, K., \& Nomoto, K. 2005, \apj, 624, 898

\bibitem[Edvardsson et al.(1993)]{edv93} 
Edvardsson, B., Andersen, J., Gustafsson, B., Lambert, D. L., Nissen, P. E.,
 \& Tomkin, J. 1993, \aap, 275, 101

\bibitem[Fran\c cois \etal (2004)]{fra04} Fran\c cois, P., Matteucci, F.,
			   Cayrel, R., Spite, M., Spite, F., \&
			   Chiappini, C. 2004, \aap, 421, 613

\bibitem[Frebel \etal (2005)]{fre05} Frebel, A., \etal\ 2005, \nat, 434, 871

\bibitem[Fr\"ohlich \etal (2006a)]{fro06}
Fr\"ohlich, C., Mart\'inez-Pinedo, G., Liebend\"orfer, M., 
Thielemann, F.-K., Bravo, E., Hix, W. R., Langanke, K.,
\& Zinner, N.T. 2006a, Phys. Rev. Let., 96, 142502

\bibitem[Fr\"ohlich \etal (2006b)]{fro04} Fr\"ohlich, C., \etal\ 2006b,
			   \apj, 637, 415

\bibitem[Fryer \etal(2006)]{fry06} Fryer, C. L., Young,
			   P. A., \& Hungerford, A. L. 2006, \apj,
			   650, 1028

\bibitem[Galama \etal(1998)]{gal98} Galama, T. J., \etal\ 1998,
\nat, 395, 670

\bibitem[Gal-Yam \etal(2004)]{gal04} Gal-Yam, A., \etal\ 2004, \apj,
			   609, L59

\bibitem[Gratton \& Sneden (1991)]{gra91}
Gratton, R. G. \& Sneden, C. 1991, \aap, 241, 501

\bibitem[Gratton et al.(2003)]{gra03}
Gratton, R. G. et al. 2003, \aap, 404, 187

\bibitem[Hachisu \etal (1990)]{hac90} Hachisu, I., Matsuda, T., Nomoto,
			   K., \& Shigeyama, T. 1990, \apj, 358, L57

\bibitem[Heger \& Langer(2000)]{heg00} Heger, A., \& Langer, N. 2000,
			   \apj, 544, 1016

\bibitem[Heger \& Woosley (2002)]{heg02} Heger, A., \& Woosley,
				S.E. 2002, \apj, 567, 532

\bibitem[Hjorth \etal(2003)]{hjo03} Hjorth, J., \etal\ 2003, \nat, 423, 847

\bibitem[Hoffman \etal(1996)]{hof96} Hoffman, R.D., Woosley, S.E.,
			   Fuller, G.M., \& Meyer, B.S. 1996, \apj, 460,
			   478

\bibitem[HO04()]{hon04} Honda, S., Aoki, W., Kajino,
			   T., Ando, H., Beers, T. C.,
			   Izumiura, H., Sadakane, K.,
			   Takada-Hidai, M. 2004, \apj, 607, 474 (HO04)

\bibitem[Imbriani \etal (2001)]{imb01} Imbriani, G., Limongi, M.,
				Gialanella, L., Terrasi, F., Straniero,
				O., \& Chieffi, A. 2001, \apj, 558, 903

\bibitem[Ishimaru \& Wanajo(1999)]{ish99} Ishimaru, Y., \& Wanajo,
			   S. 1999, \apj, 511, L33

\bibitem[Ishimaru \etal(2004)]{ish04} Ishimaru, Y., Wanajo, S., Aoki,
			   W., \& Ryan, S.G. 2004, \apj, 600, L47

\bibitem[Iwamoto \etal(1994)]{iwa94} Iwamoto, K., \etal\ 1994,
\apj, 437, L115

\bibitem[Iwamoto \etal(1998)]{iwa98} Iwamoto, K., \etal\ 1998,
\nat, 395, 672

\bibitem[Iwamoto \etal(2000)]{iwa00} Iwamoto, K., \etal\ 2000, \apj,
			   534, 660

\bibitem[IW05()]{iwa05} Iwamoto, N., Umeda, H., Tominaga,
				N., Nomoto, K., \& Maeda, K. 2005,
				Science, 309, 451 (IW05)

\bibitem[Kawabata \etal(2002)]{kaw02} Kawabata, K., \etal\ 2002, \apj,
			   580, 39

\bibitem[Kifonidis \etal (2003)]{kif03} Kifonidis, K., Plewa, T., Janka,
			   H.-Th., \& M\"uller, E. 2003, \aap, 408, 621

\bibitem[Kobayashi \etal (2006)]{kob05} Kobayashi, C., Umeda, H.,
			   Nomoto, K., Tominaga, N., \& Ohkubo, T. 2006,
			   \apj, 653, 1145

\bibitem[Kudritzki(2000)]{kud00} Kudritzki, R.-P. 2000, in The First Stars. eds. Weiss, A., Abel, T. G., \& Hill, V. (Berlin: Springer), 127

\bibitem[Langer(1992)]{lan92} Langer, N. 1992, \aap, 265, L17

\bibitem[Leonard \etal(2002)]{leo02} Leonard, D., \etal\ 2002, 
			   Pub. Astron. Soc. Pacific, 114, 1333

\bibitem[Liebend\"orfer \etal (2005)]{lie05} Liebend\"orfer, M.,
			   Rampp, M., Janka, H.-Th., \& Mezzacappa,
			   A. 2005, \apj, 620, 840

\bibitem[Limongi \& Chieffi(2003)]{lim03} Limongi, M., \& Chieffi,
			   A. 2003, \apj, 592, 404

\bibitem[Lipkin \etal(2004)]{lip04} Lipkin, Y. M., \etal\ 2004, \apj,
			   606, 381

\bibitem[MacFadyen \& Woosley(1999)]{mac99} MacFadyen, A., \& Woosley,
			   S.E. 1999, \apj, 524, 262

\bibitem[Maeda \etal(2006a)]{mae06a} Maeda, K., Mazzali, P. A., \&
			   Nomoto, K. 2006a, \apj, 645, 1331

\bibitem[Maeda \etal (2002)]{mae02} Maeda, K., Nakamura, T., Nomoto, K.,
			   Mazzali, P. A., Patat, F., \& Hachisu,
			   I. 2002, \apj, 565, 405

\bibitem[Maeda \& Nomoto(2003)]{mae03} Maeda, K., \& Nomoto, K. 2003,
                             \apj, 598, 1163

\bibitem[Maeda \etal(2006b)]{mae06b} Maeda, K., Nomoto, K., Mazzali, P. A.,
			   \& Deng, J. 2006b,
			   \apj, 640, 854

\bibitem[Maeda \& Tominaga(2007)]{mae06c} Maeda, K., \& Tominaga,
			   N. 2007, \apj, submitted

\bibitem[Maeder \& Meynet (2000)]{mae00} Maeder, A., \& Meynet, G. 2000,
			   \araa, 38, 143

\bibitem[Malesani \etal(2004)]{mal04} Malesani, D., \etal\ 2004, \apj,
			   609, L5

\bibitem[Matheson \etal(2003)]{mat03} Matheson, T., \etal\ 2003, \apj,
			   599, 394

\bibitem[Mazzali \etal(2000)]{maz00} Mazzali, P. A.,
			   Iwamoto, K., \& Nomoto, K. 2000, \apj, 545, 407

\bibitem[Mazzali \etal(2002)]{maz02} Mazzali, P. A., \etal\ 2002, \apj,
			   572, L61

\bibitem[Mazzali \etal(2003)]{maz03} Mazzali, P. A., \etal\ 2003, \apj,
			   599, L95

\bibitem[Mazzali \etal (2005)]{maz05} Mazzali, P. A., \etal\ 2005, Science, 308, 1284

\bibitem[Mazzali \etal(2006a)]{maz06} Mazzali, P. A., \etal\ 2006a, \apj,
			       645, 1323

\bibitem[Mazzali \etal(2006b)]{maz06b} Mazzali, P. A., \etal\ 2006b, \nat,
			       442, 1018

\bibitem[Meynet \& Maeder (2005)]{mey05} Meynet, G., \& Maeder, A. 2005,
				\aap, 429, 581

\bibitem[McWilliam (1997)]{mcw97} McWilliam, A. 1997, \araa,
				35, 503

\bibitem[McWilliam (1998)]{mcw98} McWilliam, A. 1998, \aj,
				115, 1640

\bibitem[McWilliam \etal (1995a)]{mcw95a} McWilliam, A., Preston, G. W.,
				Sneden, C., \& Searle, L. 1995a, \aj,
				109, 2757

\bibitem[McWilliam \etal (1995b)]{mcw95b} McWilliam, A., Preston, G. W.,
				Sneden, C., \& Shectman, S. 1995b, \aj,
				109, 2736

\bibitem[Nagataki \etal(2006)]{nag06} Nagataki, S., Mizuta, A., \& Sato,
			   K. 2006, \apj, 647, 1255

\bibitem[Nagataki \etal(2003)]{nag03} Nagataki, S., Mizuta, A., Yamada,
			   S., Takabe, H., \& Sato, K. 2003, \apj, 596, 401

\bibitem[Nakamura \& Umemura (1999)]{nakf99} Nakamura, F., \& Umemura,
				M. 1999, \apj, 515, 239

\bibitem[Nakamura \etal (2001a)]{nak01} Nakamura, T., Mazzali, P.A., Nomoto,
				K., \& Iwamoto, K. 2001a, \apj, 550, 991

\bibitem[Nakamura \etal (1999)]{nak99} Nakamura, T., Umeda, H., Nomoto,
				K., Thielemann, F.-K., \& Burrows,
				A. 1999, \apj, 517, 193

\bibitem[Nakamura \etal (2001b)]{nak01b} Nakamura, T., \etal\ 2001b,
			   \apj, 555, 880

\bibitem[Nissen \etal(2002)]{nis02} Nissen, P.E., Primas, F., Asplund,
			   M., \& Lambert, D.L. 2002, \aap, 390, 235

\bibitem[Nomoto \& Hashimoto (1988)]{nom88} Nomoto, K., \& Hashimoto,
			   M. 1988, \physrep, 163, 13

\bibitem[Nomoto \etal (1997)]{nom97} Nomoto, K., Hashimoto, M.,
				Tsujimoto, T., Thielemann, F.-K.,
				Kishimoto, M., Kubo, Y., \& Nakasato,
				N. 1997, \nphysa, 616, 79

\bibitem[Nomoto \etal (2004a)]{nom03} Nomoto, K., Maeda, K., Mazzali,
			   P. A., Umeda, H., Deng, J., \& Iwamoto,
			   K. 2004a, in Stellar Collapse, ed. C. L. Fryer
			   (Dordrecht: Kluwer), 277
			    (astro-ph/0308136)

\bibitem[Nomoto \etal (2004b)]{nom04} Nomoto, K., Maeda, K., Umeda, H.,
				Tominaga, N., Ohkubo, T., Deng, J., \&
				Mazzali, P.A. 2004b, MmSAI, 75, 312

\bibitem[Nomoto \etal (2001)]{nom01}
Nomoto, K., Mazzali, P.A., Nakamura, T., et al. 2001, in Supernovae and
			   Gamma Ray Bursts, ed. M. Livio et
			   al.
			   (Cambridge: Cambridge Univ. Press), 144 (astro-ph/0003077)

\bibitem[Nomoto \etal (1993)]{nom93} Nomoto, K., Suzuki, T., Shigeyama,
			   T., Kumagai, S., Yamaoka, H., \& Saio,
			   H. 1993, \nat, 364, 507

\bibitem[Nomoto \etal (2006)]{nom06} Nomoto, K., Tominaga, N., Umeda, H.,
				Kobayashi, C., \& Maeda, K. 2006,
			   \nphysa, 777, 424 (astro-ph/0605725)

\bibitem[Nomoto \etal (1994)]{nom94} Nomoto, K., Yamaoka, H., Pols,
			   O. R., van den Heuvel, E. P. J., Iwamoto, K.,
			   Kumagai, S., \& Shigeyama,
			   T. 1994, \nat, 371, 227

\bibitem[Norris \etal(2001)]{nor01} Norris, J. E., Ryan,
			   S. G., \& Beers, T. C. 2001, \apj, 561, 1034

\bibitem[Ohkubo \etal (2006)]{ohk05} Ohkubo, T., Umeda, H., Maeda, K.,
			   Nomoto, K., Tsuruta, S., \& Rees, M. J. 2006,
			   \apj, 645, 1352

\bibitem[Patat \etal(2001)]{pat01} Patat, F., \etal\ 2001, \apj, 555, 900

\bibitem[Pian \etal(2006)]{pia06} Pian, E., \etal\ 2006, \nat, 442, 1011

\bibitem[Podsiadlowski \etal(2004)]{pod04} Podsiadlowski, Ph., Mazzali,
			   P. A., Nomoto, K., Lazzati, D., \& Cappellaro,
			   E. 2004, \apj, 607, L17

\bibitem[Primas et al.(2000)]{pri00}
Primas, F., Reimers, D., Wisotzki, L., Reetz, J., Gehren, T., \& Beers,
			   T. C. 2000, in The First Stars, ed. A. Weiss,
			   T. Abel, \&, V. Hill (Berlin: Springer), 51

\bibitem[Pruet \etal(2004a)]{pru04} Pruet, J., Surman, R.,
			   \& McLaughlin, G.C. 2004a, \apj, 602, L101

\bibitem[Pruet \etal(2004b)]{pru04b} Pruet, J.,
			   Thompson, T.A., \& Hoffman, R.D. 2004b, \apj,
			   606, 1006

\bibitem[Pruet \etal (2005)]{pru05} Pruet, J., Woosley, S. E., Buras, R.,
				Janka, H.-T., \& Hoffman, R. D. 2005,
				\apj, 623, 325

\bibitem[Rampp \& Janka (2000)]{ram00} Rampp, M., \& Janka, H.-Th. 2000,
			    \apj, 539, L33

\bibitem[Ryan \etal(2005)]{rya05} Ryan, S. G., Aoki, W., Norris, J. E.,
			     \& Beers, T. C. 2005, \apj, 635, 349

\bibitem[Ryan \etal (1996)]{rya96} Ryan, S. G., Norris, J. E.,
			     \& Beers, T. C. 1996, \apj, 471, 254 

\bibitem[Sauer \etal(2006)]{sau06} Sauer, D. N., Mazzali, P. A., Deng,
			   J., Valenti, S., Nomoto, K., \& Filippenko,
			   A. V. 2006, \mnras, 369, 1939

\bibitem[Shigeyama \& Nomoto (1990)]{shi90} Shigeyama, T., \&
                             Nomoto, K. 1990, \apj, 360, 242

\bibitem[Shigeyama \etal(1994)]{shi94} Shigeyama, T., Suzuki, T.,
			   Kumagai, S., Nomoto, K., Saio,
			   H., \& Yamaoka, H. 1994, \apj, 420, 341

\bibitem[Shigeyama \& Tsujimoto (1998)]{shi98} Shigeyama, T., \&
                             Tsujimoto, T. 1998, \apj, 507, L135

\bibitem[Sneden et al.(1991)]{sne91}
Sneden, C., Gratton, R. G., \& Crocker, D. A. 1991, \aap, 246, 354

\bibitem[Spergel \etal(2003)]{spe03} Spergel, D. N., \etal\ 2003, \apjs,
			   148, 175

\bibitem[Spite \etal(2005)]{spi05} Spite, M., \etal\ 2005, \aap, 430, 655

\bibitem[Spite \etal(2006)]{spi06} Spite, M., \etal\ 2006, \aap, 455, 291

\bibitem[Stanek \etal(2003)]{sta03} Stanek, K. Z., \etal\ 2003, \apj,
			   591, L17

\bibitem[Suda \etal(2004)]{sud04} Suda, T., Aikawa, M., Machida, M. N.,
			   Fujimoto, M. Y., \& Iben, I., Jr. 2004, \apj,
			   611, 476

\bibitem[Taubenberger \etal(2006)]{tau06} Taubenberger, S., \etal\ 2006,
			   \mnras, 371, 1459

\bibitem[Thielemann \etal(1996)]{thi96} Thielemann,
			   F.K., Nomoto, K., \& Hashimoto, M. 1996,
			   \apj, 460, 408

\bibitem[Thomsen \etal(2004)]{tho04} Thomsen, B., \etal\ 2004, \aap,
			   419, L21

\bibitem[Thornton \etal(1998)]{tho98} Thornton, K., Gaudlitz, M., Janka,
			   H.-Th., \& Steinmetz, M. 1998, \apj, 500, 95

\bibitem[Tominaga \etal(2007)]{tom06} Tominaga, N., Maeda, K., Umeda,
			   H., Nomoto, K., Tanaka, M., Iwamoto, N., \&
			   Mazzali, P.A. 2007, \apj, submitted

\bibitem[Tominaga \etal(2005)]{tom05} Tominaga, N., \etal\ 2005, \apj,
			   633, L97

\bibitem[Tomita \etal(2006)]{tomita06} Tomita, H., \etal\ 2006, \apj,
			   644, 400

\bibitem[Turatto \etal(1998)]{tur98} Turatto, M., \etal\ 1998, 498, L129

\bibitem[Tumlinson (2006)]{tum05} Tumlinson, J. 2006, \apj, 641, 1

\bibitem[UN02a()]{ume02} Umeda, H., \& Nomoto, K. 2002a,
\apj, 565, 385 (UN02a)

\bibitem[UN03()]{ume03} Umeda, H., \& Nomoto, K. 2003,
\nat, 422, 871 (UN03)

\bibitem[UN05()]{ume05} Umeda, H., \& Nomoto, K. 2005,
\apj, 619, 427 (UN05)

\bibitem[UNN00()]{ume00} Umeda, H., Nomoto,
			     K., \& Nakamura, T. 2000, in The First
			   Stars. eds. Weiss, A., Abel, T. G., \& Hill,
			   V. (Berlin: Springer), 150 
			   (astro-ph/9912248) (UNN00)

\bibitem[Umeda \etal(2002b)]{ume02b}
Umeda, H., Nomoto, K., Tsuru, T., \& Matsumoto, H. 2002b, \apj, 578, 855

\bibitem[Wanajo (2006a)]{wan06a} Wanajo, S. 2006a, \apj, 647, 1323

\bibitem[Wanajo (2006b)]{wan06b} Wanajo, S. 2006b, \apj, 650, L79

\bibitem[Wang \etal (2002)]{wan02} Wang, L., \etal\ 2002, \apj, 579, 671

\bibitem[Wang \etal(2003)]{wan03} Wang, L., \etal\ 2003, \apj, 592, 457

\bibitem[Wasserburg \& Qian (2000)]{was00} Wasserburg, G. J., \& Qian,
				Y.-Z. 2000, \apj, 529, L21

\bibitem[Weiss \etal(2004)]{wei04} Weiss, A., Schlattl, H., Salaris, M.,
			   \& Cassisi, S. 2004, \aap, 422, 217

\bibitem[Woosley \etal(1999)]{woo99} Woosley, S. E.,
			   Eastman, R.G., \& Schmidt, B.P. 1999, \apj,
			   516, 788

\bibitem[Woosley \& Weaver (1993)]{woo93} Woosley, S. E., \& Weaver,
			     T. A. 1993, \physrep, 227, 65


\bibitem[WW95()]{woo95} Woosley, S. E., \& Weaver,
			     T. A. 1995, \apjs, 101, 181 (WW95)

\bibitem[Zampieri \etal(2003)]{zam03} Zampieri, L., Pastorello, A.,
			   Turatto, M., Cappellaro, E., Benetti, S.,
			   Altavilla, G., Mazzali, P., \& Hamuy, M. 2003,
			   \mnras, 338, 711

\end{thebibliography}
\end{document}